\documentclass{aa}  

\usepackage{graphicx}
\usepackage{txfonts}
\usepackage[colorlinks]{hyperref}
\usepackage{bm}
\usepackage{makecell}

\usepackage{textcomp}
\usepackage{gensymb}
\usepackage{xcolor}
\usepackage{soul}
\newcommand{\orcidlink}[1]{\href{https://orcid.org/#1}{\includegraphics[width=10pt]{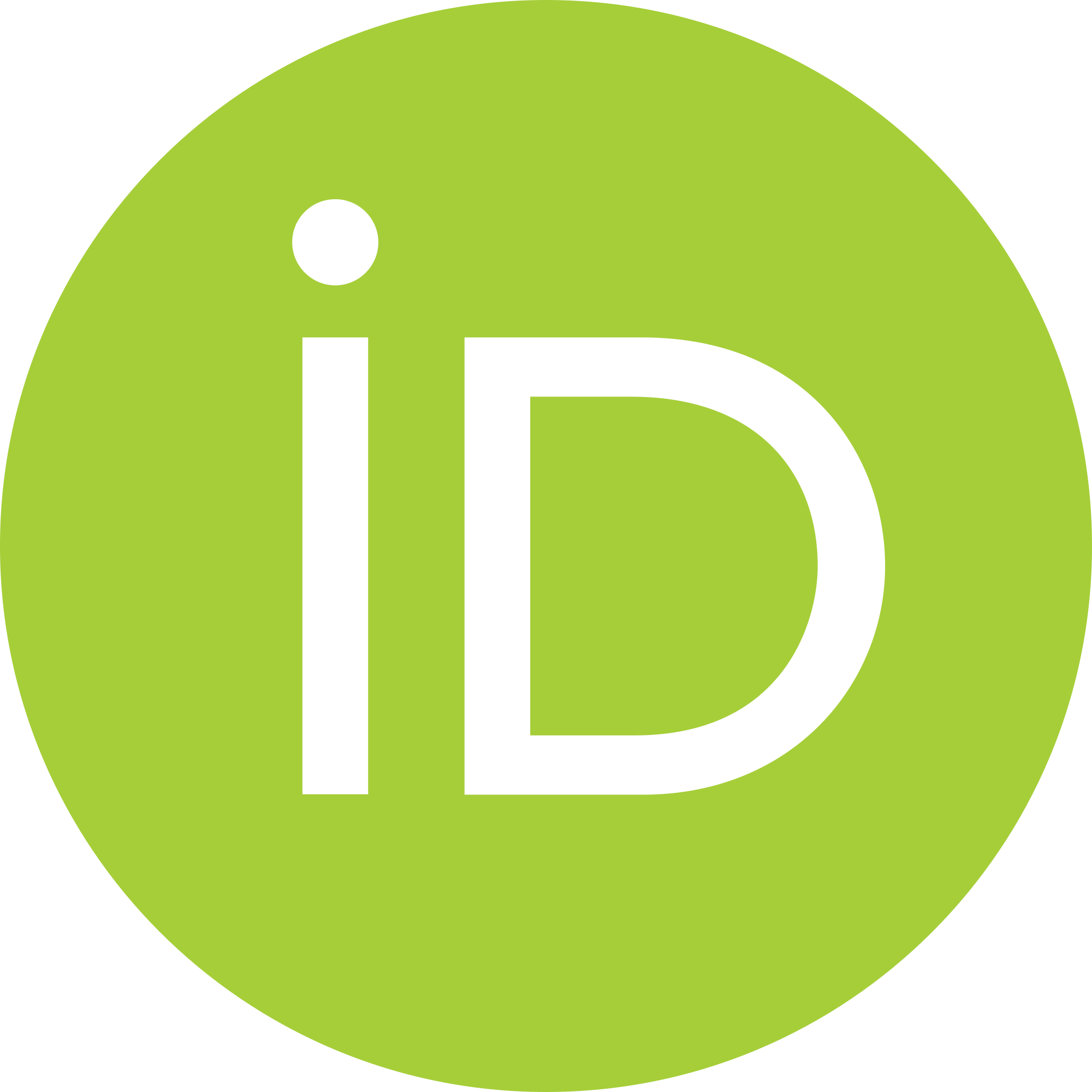}}}

\usepackage[most]{tcolorbox}
\definecolor{light-gray}{gray}{0.95}
\tcbset{on line, 
        boxsep=.7pt, left=.3pt, right=.3pt, top=0pt, bottom=0pt,
        colframe=light-gray, colback=light-gray,  
        highlight math style={enhanced},
        }

\newcommand{\ssim}{\mathchar"5218\relax\,}
\renewcommand{\Re}{{\rm Re}}

\definecolor{linkcolor}{rgb}{0.0,0.3,0.5}
 \hypersetup{
     colorlinks=true,
     linkcolor=linkcolor,
     filecolor=linkcolor,
     citecolor = linkcolor,      
     urlcolor=linkcolor,
     }

\def\apjl{{Astrophys.~J.~Lett.}}
\def\apjs{{Astrophys.~J.~Supp.}}
\def\ao{{Appl.~Opt.}}

\def\aap{{Astron.~Astrophys.}}

\defcitealias{giacobbo2020}{GM20}
\defcitealias{kroupa2001}{K01}
\defcitealias{larson1998}{L98}
\defcitealias{sana2012}{S12}
\defcitealias{stacy2013}{SB13}
\defcitealias{hartwig2022}{H22}
\defcitealias{jaacks2019}{J19}
\defcitealias{liu_bromm_20}{LB20}
\defcitealias{skinner2020}{SW20}
\defcitealias{madau2017}{MF17}
\defcitealias{santoliquido2023}{S23}

\makeatletter
\renewcommand*\aa@numarticle{}
\renewcommand*\aa@textidlineempty{\aa@headfont \aa@headings}
\renewcommand*\aa@pageof{page \thepage{} of \pageref*{LastPage}}
\renewcommand*\aa@journalname{Accepted in Astronomy \& Astrophysics}
\renewcommand*\aa@manuscriptname{%
  \hspace{\stretch{1}}%
  \copyright ESO \the\year
}
\makeatother
\usepackage{letltxmacro}
\LetLtxMacro{\originaleqref}{\eqref}
\renewcommand{\eqref}{Eq.~\originaleqref}
\addto\extrasenglish{%
}

\begin{document}

   \title{Classifying binary black holes from Population III stars with the \textit{Einstein} Telescope: A machine-learning approach}
   
   \author{Filippo Santoliquido\thanks{Corresponding author; \href{mailto:filippo.santoliquido@gssi.it}{filippo.santoliquido@gssi.it}}$\,$\orcidlink{0000-0003-3752-1400}\inst{1,2},
   Ulyana Dupletsa$\,$\orcidlink{0000-0003-2766-247X}\inst{1,2},
   Jacopo Tissino$\,$\orcidlink{0000-0003-2483-6710}\inst{1,2},
Marica Branchesi$\,$\orcidlink{0000-0003-1643-0526}\inst{1,2},\\
Francesco Iacovelli$\,$\orcidlink{0000-0002-4875-5862}\inst{3,4},
Giuliano Iorio$\,$\orcidlink{0000-0003-0293-503X}\inst{5,6},
Michela Mapelli$\,$\orcidlink{0000-0001-8799-2548}\inst{7,5,6},
Davide Gerosa$\,$\orcidlink{0000-0002-0933-3579}\inst{8,9,10},
Jan Harms$\,$\orcidlink{0000-0002-7332-9806}\inst{1,2},\\
\and Mario Pasquato$\,$\orcidlink{0000-0003-3784-5245}\inst{5,6,11,12,13} 
}
   \institute{Gran Sasso Science Institute (GSSI), 67100 L’Aquila, Italy
    \and
    INFN, Laboratori Nazionali del Gran Sasso, 67100 Assergi, Italy
\and
D\'epartement de Physique Th\'eorique, Universit\'e de Gen\`eve, 24 quai Ernest Ansermet, 1211 Gen\`eve, Switzerland
\and
Gravitational Wave Science Center (GWSC), Universit\'e de Gen\`eve, 1211 Gen\`eve, Switzerland
    \and
    Dipartimento di Fisica e Astronomia ``G. Galilei'', Università degli studi di Padova, Vicolo dell’Osservatorio 3, 35122 Padova,
Italy
\and 
INFN, Sezione di Padova, via Marzolo 8, 35131 Padova, Italy
\and
Institut für Theoretische Astrophysik, ZAH, Universität Heidelberg, Albert-Ueberle-Str. 2, 69120 Heidelberg, Germany
    \and Dipartimento di Fisica ``G. Occhialini'', Università degli studi di Milano-Bicocca, piazza della Scienza 3, 20126 Milano, Italy
\and
INFN, Sezione di Milano-Bicocca, piazza della Scienza 3, 20126 Milano, Italy
\and
School of Physics and Astronomy \& Institute for Gravitational Wave Astronomy, University of Birmingham, Birmingham, B15 2TT, United Kingdom
\and
D\'{e}partement de Physique, Universit\'{e} de Montr\'{e}al, 1375 Avenue Th\'{e}r\`{e}se-Lavoie-Roux, Montr\'{e}al, Canada
\and 
Mila -- Quebec Artificial Intelligence Institute, 6666 Rue Saint-Urbain, Montr\'{e}al, Canada
\and
Ciela -- Montr\'{e}al Institute for Astrophysical Data Analysis and Machine Learning, Montr\'{e}al, Canada
             }


\abstract{Third-generation (3G) gravitational-wave detectors such as the \textit{Einstein} Telescope (ET) will observe binary black hole (BBH) mergers at redshifts up to $z\sim 100$. However, an unequivocal determination of the origin of high-redshift sources will remain uncertain because of the low signal-to-noise ratio (S/N) and poor estimate of their luminosity distance. This study proposes a machine-learning approach to infer the origins of high-redshift BBHs. We specifically differentiate those arising from Population III (Pop.~III) stars, which probably are the first progenitors of star-born BBH mergers in the Universe, and those originated from Population I-II (Pop.~I-II) stars. We considered a wide range of models that encompass the current uncertainties on Pop.~III BBH mergers. We then estimated the parameter errors of the detected sources with ET using the Fisher information-matrix formalism, followed by a classification using \textsc{XGBoost}, which is a machine-learning algorithm based on decision trees. For a set of mock observed BBHs, we provide  the probability that they belong to the Pop.~III class while considering the parameter errors of each source. In our fiducial model, we accurately identify $\gtrsim 10\%$ of the detected BBHs that originate from Pop.~III stars with a precision $>90\%$. Our study demonstrates that machine-learning enables us to achieve some pivotal aspects of the ET science case by exploring the origin of individual high-redshift GW observations. We set the basis for further studies, which will integrate additional simulated populations and account for further uncertainties in the population modeling.}


\titlerunning{Classifying binary black holes from Population III stars with the \textit{Einstein} Telescope}
\authorrunning{Santoliquido et al.}

\maketitle
%

\section{Introduction}
The \textit{Einstein} Telescope (ET) will be a forefront instrument for observing gravitational waves (GWs) in Europe \citep{punturo2010,maggiore2020}. In particular, the ET will observe mergers of binary black holes (BBHs) at redshifts as high as $z \sim 100$ \citep{ng2021,kalogera2021,ng2022,2022A&A...663A.156Y,branchesi2023}. However, the detection of high-redshift events does not automatically imply an accurate inference of the their parameters. Several works  have demonstrated the challenge the ET will face in 
constraining the redshift distribution of the detected sources, which in turn is a crucial step to unveil their astrophysical origins \citep[e.g.][]{Ng2022b,chen22,iacovelli2022b,mancarella2023,Fairhurst2023,marcoccia2023}.

To this end, \cite{Ng2022b,Ng2023}  investigated the accuracy of redshift measurements for BBH mergers. They specifically sought to determine whether their primordial origin could be established. Their findings indicate that for BBHs with total masses ranging between 20~M$_\odot$ and 40~M$_\odot$ that merge at $z > 40$, it is possible to infer that $z > 30$  with a credibility of up to 70\% using a single ET observatory. \cite{mancarella2023} studied a population of BBHs with parameter distributions extrapolated from those of current LIGO-Virgo-KAGRA events \citep{O3b} and found that only $\sim 3\%$ of the sources with $z \geq 30$ can be reliably constrained to be beyond redshift $30$ at a credibility of 99\%. These findings limit the scientific potential of the ET and Cosmic Explorer \citep{reitze2019} because the detection of BBHs in the early Universe is a pivotal element of the science objectives for third-generation (3G) GW interferometers.

Population III (Pop.~III) stars  form from pristine metal-free gas follow cosmic nucleosynthesis at $z > 20$ \citep{Haiman1996,Tegmark1997,abel2002,Yoshida2003,klessen2023} and are thought to lead to the formation of the first star-origin black holes in the Universe \citep{kinugawa2014,kinugawa2016,Hartwig2016,belczynski2017,tanikawa2022,costa2023,Nandal2023,iwaya2023,tanikawa2024,Boyuan2024,liu2023arXiv}. There have been no direct observations of Pop.~III stars so far \citep{rydberg2013,Schauer2022,Larkin2023,meena2023, Trussler2023}, and they can directly be detected with the \textit{James Webb} Space Telescope (JWST) only with gravitational lensing \citep{Zackrisson2023,bovill2024,Wiggins2024}. In recent years, BBHs originating from Pop.~III stars have gained significant attention in the context of GW astronomy \citep{kinugawa2020,liubromm2020,liu_bromm_20,tanikawa2022,tanikawa2022b,wang2022}. In particular, \cite{santoliquido2023} showed that between $\ssim 20\%$ and $\ssim 70\%$ of the detectable mergers from Pop.~III BBHs occur at $z > 8$. These percentages depend on assumptions regarding the star formation rate density 
and the adopted initial conditions, resulting in distinct redshift distributions. 

We present a novel method for distinguishing the origins of GW sources. Specifically, we differentiate between BBHs originating from Pop.~III stars and those from Pop.~I-II stars, which encompass sources that merge across a broad redshift range. It can be highly informative to determine the origin of single GW events, especially regarding the specific physical processes leading to their formation \citep{abbottO2popandrate,Fishbach2020,Kimball2020,singh2022,tong2022,gamba2023,Ng2023}. Two distinct methods have been proposed in the literature to achieve this. The first method involves a classification in which categorical variables are inferred through a hierarchical Bayesian analysis that considers hyperparameters describing each subpopulation \citep{farr2015,mould2022,mould2023, Godfrey2023}. The second method is based on machine-learning (ML), where ML algorithms are trained on simulated sources and are subsequently applied to GW detections \citep{antonelli2023}. These methods have both been directly applied to GW events published in LIGO-Virgo-KAGRA catalogs \citep{abbottGW150914,2019PhRvX...9c1040A,2021PhRvX..11b1053A,O3b}. 

In our analysis, we combine the outcomes of ML classifiers trained on Pop.~III and Pop.~I-II BBHs with the inferred posterior distribution of detected sources and take the parameter estimation capabilities of the ET into account. Consequently, we assign a probability for each source to be associated with either origin. We assess the performance of our classification and determine how the data inform the classification process. Our investigation covers systematic variations within various models describing Pop.~III BBHs, including astrophysical merger rates.

With our proposed method, we aim to demonstrate that it is 
possible to infer the origin of individual events, considering the limited but still significant capability of the ET to perform the parameter estimation of high-redshift sources \citep{Ng2022b,Ng2023,mancarella2023}. Our method combines information from astrophysical simulations with the results of the parameter estimation to achieve this goal. It is complementary to currently adopted methods that are based on Bayesian analysis \citep[e.g.,][]{thrane2019, mould2023}. The focus of our method on individual GW observations potentially enables us to identify rare events such as the most distant Pop.~III BBHs, which are particularly interesting for constraining our understanding of the early Universe.

Machine-learning, deep learning, and artificial intelligence have become increasingly critical in astronomy and astrophysics \citep[e.g.,][]{astroMLText,baron2019,Djorgovski2022,Moriwaki2023,pasquato2023,riggi2024}, with a growing emphasis on applications within the GW field \citep{zevin2017,cuoco2021b,cuoco2021a}. In this context, a significant focus of ML techniques involves classification tasks \citep[e.g.,][]{baker2015,Chatterjee2020,sasaoka2023,Alhassan2023,antonelli2023,berbel2023}. Furthermore, the large number of detections expected with 3G GW observatories  \citep{reitze2019,maggiore2020,branchesi2023} will make a full Bayesian parameter estimation computationally expensive \citep{Couvares2021,Roulet2024}. Hence, ML and deep learning hold particular promise for a rapid inference of GW parameters \citep{green2020,dax2021,dax2023,Williams2021,Gabbard2022,Wong2023,Wouters2024}.

This paper is organized as follows: We delineate in Sect.~\ref{sec:astro} the simulation setup for generating BBHs from Pop.~III and Pop.~I-II stars. In Sect.~\ref{sec:cosmorate} we evaluate the merger rate density as a function of redshift, while Sect.~\ref{sec:pe} provides a brief overview of the Fisher information-matrix (FIM) approximation we used to estimate the parameters. Section~\ref{sec:post_samples} details the method for assigning the probability of each source to belong to the Pop.~III class, while Sect.~\ref{sec:classifier} delineates the training process for the ML classifiers. Our results along with a discussion of our method caveats are presented in Sects.~\ref{sec:results} and \ref{sec:discussion}. Conclusion are drawn in Sect.~\ref{sec:conclusion}.

\section{Methods}

\subsection{Astrophysical populations}
\label{sec:astro}

\cite{santoliquido2023} presented a comprehensive investigation on the main sources of uncertainty affecting BBHs born from Pop.~III stars. Here, we focus on three of their models: their fiducial model, along with the two models yielding the lowest and highest comoving merger rate density of those they calculated.

\begin{table*}
    \caption{Astrophysical populations.}  
    \begin{center}
        \begin{tabular}{lllllllr}
            \hline\hline
            Name & IC &  $M_{\rm ZAMS,1}$   &  $q$     & $P$      &  $e$ & SFRD & $\mathcal{N}$ 
            \\  
            \hline
            Pop.~III fiducial &   LOG1   & \eqref{eq:loguniform}  & \eqref{eq:sanaq}         & \eqref{eq:sanap}       & \eqref{eq:sanae}  & \citetalias{hartwig2022} & 6916  
            \\
            
            Pop.~III optimistic & LAR1     & \eqref{eq:larson}  & \eqref{eq:sanaq}        & \eqref{eq:sanap}          & \eqref{eq:sanae}  & \citetalias{jaacks2019} & 119399 
            \\
            
            Pop.~III pessimistic & TOP5     & \eqref{eq:top} & \eqref{eq:stacyq}          & \eqref{eq:stacyp}          &  \eqref{eq:thermal}  & \citetalias{skinner2020}& 202  
            \\
                        Pop.~I-II & Fiducial     & \eqref{eq:kroupa} & \eqref{eq:sanaq}          & \eqref{eq:sanap}          & \eqref{eq:sanae}  & \citetalias{madau2017} & 887364 
                        \\
            \hline
        \end{tabular}
    \end{center}  
    \label{tab:IC} 
    \tablefoot{For each model, we list the name of the initial condition configuration (IC), the distribution of the ZAMS masses of the primary star $M_{\rm ZAMS,1}$, the distribution of the mass ratios $q$, the distribution of the orbital periods $P$, the distribution of the orbital eccentricities $e$, and the adopted SFRD model. The rightmost column reports the total number of mergers from \eqref{eq:n_sources} assuming $T_{\mathrm{obs}} = 10$~yr.}
\end{table*}

\subsubsection{Population III stars} 

The  catalogs of BBH mergers reported by \cite{santoliquido2023} were generated using the binary population synthesis code \textsc{sevn} \citep{spera2019,mapelli2020} for both Pop.~III \citep{costa2023} and  Pop.~I-II stars \citep{iorio2023}. \textsc{sevn} combines single and binary evolution by interpolating a set of precomputed single stellar-evolution tracks \citep{iorio2023}. Pop.~III stellar tracks were computed using the \textsc{parsec} code \citep{bressan2012,costa2021,nguyen2022} with a metallicity of $Z = 10^{-11}$ and covering a zero-age main-sequence (ZAMS) mass range of 2.2-600~M$_\odot$ (for details, see \citealt{costa2023}).

\cite{santoliquido2023} explored a large set of initial conditions for Pop.~III binary systems. The three models analyzed here correspond to the setups labeled  LOG1, LAR1, and TOP5. The initial mass function (IMF) is defined as a flat-in-log probability distribution for model LOG1 \citep{stacy2013,susa2014, hirano2015,wollenberg2020,chon2021,tanikawa2021,jaura2022,prole2022},
\begin{equation}
\label{eq:loguniform}
    \xi(M_{\mathrm{ZAMS,1}}) \propto M_{\mathrm{ZAMS,1}}^{-1},
\end{equation}
a \cite{larson1998} distribution  for  model LAR1,
\begin{equation}
\label{eq:larson}
    \xi(M_{\mathrm{ZAMS,1}}) \propto M_{\mathrm{ZAMS,1}}^{-2.35}e^{M_\mathrm{cut}/M_\mathrm{ZAMS,1}}
\end{equation}
with $M_\mathrm{cut} = 20$ M$_\odot$ \citep{Valiante2016}, and a top-heavy distribution for model TOP5 \citep{stacy2013,jaacks2019,liubromm2020},
\begin{equation}
\label{eq:top}
    \xi(M_{\mathrm{ZAMS,1}}) \propto M_{\mathrm{ZAMS,1}}^{-0.17}e^{(M_\mathrm{cut}/M_\mathrm{ZAMS,1})^2}\,,
\end{equation}
with $M_\mathrm{cut} = 20$ M$_\odot$.

For the LOG1 and LAR1 models, the mass ratio ($q = M_{\mathrm{ZAMS},2}/M_{\mathrm{ZAMS},1}$), orbital period ($P$), and eccentricity ($e$) were drawn from the distributions by \cite{sana2012}. These distributions are fits to O- and B-type binary stars in the local Universe,
\begin{equation}
\label{eq:sanaq}
    \xi(q) \propto q^{-0.1}~\mathrm{with}~q\in [0.1,1]~\mathrm{and}~M_{\mathrm{ZAMS},2}>2.2~M_\odot\,,
\end{equation}
\begin{equation}
\label{eq:sanap}
    \xi(\log P) \propto (\log P)^{-0.55}~\mathrm{with}~\log P \in [0.15,5.5]\,,
\end{equation}
\begin{equation}
\label{eq:sanae}
    \xi(e) \propto e^{-0.142}~\mathrm{with}~e\in (0,1]\,.
\end{equation}

Model TOP5 instead adopts the mass ratio and the eccentricity distributions derived from cosmological simulations by \cite{stacy2013},
\begin{equation}
\label{eq:stacyq}
    \xi(q) \propto q^{-0.55}~\mathrm{with}~q\in [0.1,1]~\mathrm{and}~M_{\mathrm{ZAMS},2}>2.2~M_\odot\,,
\end{equation}
\begin{equation}
\label{eq:stacyp}
    \xi(\log P) \propto \exp [-(\log P-\mu)^2/(2\sigma^2)]\,,
\end{equation}
with $\mu = 5.5$ and $\sigma = 0.85$, and a thermal distribution for the eccentricity \citep{kinugawa2014,Hartwig2016,tanikawa2021},
\begin{equation}
\label{eq:thermal}
    \xi(e) \propto 2e~\mathrm{with}~e \in [0,1).
\end{equation}

We considered three independent estimates of the Pop.~III star formation rate density (SFRD). These were based on the models by \citet[][hereafter \citetalias{hartwig2022}]{hartwig2022}, \citet[][hereafter \citetalias{jaacks2019}]{jaacks2019}, and \citet[][hereafter \citetalias{skinner2020}]{skinner2020} and were adopted in model LOG1, LAR1 and TOP5, respectively.

We call the LOG1, LAR1, and TOP5 models fiducial, optimistic, and pessimistic, respectively. A summary of our models is reported in Table~\ref{tab:IC}.

\subsubsection{Population I-II stars}

We considered the fiducial model from \cite{iorio2023} as representative of BBHs that formed from Pop.~I-II stars. The initial ZAMS mass of primary stars follows a \cite{kroupa2001} IMF,
\begin{equation}
\label{eq:kroupa}
    \xi(M_\mathrm{ZAMS,1}) \propto M_\mathrm{ZAMS}^{-2.3}~~\mathrm{with}~M_\mathrm{ZAMS,1}\in[5,150]~\mathrm{M}_\odot\,.
\end{equation}
The secondary masses, initial orbital periods, and eccentricities were distributed using the prescriptions by \cite{sana2012} (\eqref{eq:sanaq},~(\ref{eq:sanap}), and (\ref{eq:sanae})).

\subsubsection{Population synthesis with \texorpdfstring{\textsc{sevn}}{sevn}}

We adopted the rapid model for core-collapse supernovae \citep{fryer2012} to convert the final properties of stars into BH masses. Additionally, we incorporated the outcomes of electron-capture supernovae, as described by \cite{giacobbo2019}. For pulsational pair-instability and pair-instability supernovae, we adopted the model by \cite{mapelli2020}. The black-hole natal kicks were drawn from the formalism presented by \cite{giacobbo2020}.

\textsc{sevn} integrates wind mass transfer \citep{bondi1944}, stable Roche-lobe overflow \citep{lubow1975,ulrich1976}, common-envelope evolution \citep{webbink1984}, and GW emission leading to orbital
decay and circularization \citep{peters1964}. We used the same setup as in the fiducial model by \cite{iorio2023}, using their default values for all relevant parameters. 
We adopted a common-envelope efficiency parameter of $\alpha{}=1$, which corresponds to assuming that all the orbital energy that is lost from the system contributes to unbinding the common envelope.

\subsection{\texorpdfstring{\textsc{cosmo$\mathcal{R}$ate}}{cosmoRate}}
\label{sec:cosmorate}

We determined the evolution of BBH merger rate density using \textsc{cosmo$\mathcal{R}$ate} \citep{santoliquido2020, santoliquido2021}, which interfaces catalogs of simulated BBHs with a metallicity-dependent SFRD model. The merger rate density in the source frame is given by
\begin{equation}
\label{eq:mrd}
    \mathcal{R}(z) = \int_{z_{{\rm{max}}}}^{z}\left[\int_{Z_{{\rm{min}}}}^{Z_{{\rm{max}}}} \,{}\psi(z')\,{}p(Z|z')\,{} 
    \mathcal{F}(z',z,Z) \,{}{\rm{d}}Z\right]\,{} \frac{{{\rm d}t(z')}}{{\rm{d}}z'}\,{}{\rm{d}}z'\,,
\end{equation}
where $\psi(z')$ represents the chosen SFRD evolution, selected from those presented in Table~\ref{tab:IC}, and $p(Z|z')$ is the distribution of metallicity $Z$ at a fixed formation redshift $z'$. Considering that we modeled Pop.~III stars with a single metallicity value, we simply set $p(Z|z')$ equal to a Dirac delta function centered on the chosen value  $Z=10^{-11}$. For the case of Pop.~I-II stars, we used \citep{madau2017}
\begin{equation}
\psi{}(z)= a{}\frac{(1+z)^{b}}{1+[(1+z)/c]^{d}} \,,
\label{eq:madau2017}
\end{equation}
with $a=0.01$~M$_\odot$\,Mpc$^{-3}$\,yr$^{-1}$, $b = 2.6$, $c = 3.2$, and $d = 6.2$, while for the metallicity distribution, we adopted  
  \citep{madau2017}
\begin{equation}
\label{eq:pdf2}
p(Z|z') = \frac{1}{\sqrt{2 \pi\,{}\sigma_{\rm Z}^2}}\,{} \exp\left\{{-\,{} \frac{\left[\log{(Z(z')/{\rm Z}_\odot)} - \langle{}\log{Z(z')/Z_\odot}\rangle{}\right]^2}{2\,{}\sigma_{\rm Z}^2}}\right\}\,,
\end{equation}
where $\langle{}\log{Z(z')/Z_\odot}\rangle{}=\log{\langle{}Z(z')/Z_\odot\rangle{}}-\ln{(10)}{}\sigma_{\rm Z}^2/2$ and $\sigma_Z = 0.2$ \citep{bouffanais2021b}.

In \eqref{eq:mrd} we introduced the expression ${\rm{d}}t(z')/{\rm{d}}z' = (1+z')^{-1}H(z')^{-1}$, where the Hubble parameter $H(z)$ in a flat $\Lambda$CDM Universe is given by
\begin{equation}
\label{eq:cosmo}
H(z) = H_0\sqrt{(1+z)^3\Omega_M+(1-\Omega_M)}\,,
\end{equation}
with $H_0$ denoting the Hubble constant, and $\Omega_M$ representing the a-dimensional matter density parameter. 
We used the cosmological parameters reported by \cite{Planck2018}. The expression $\mathcal{F}(z',z, Z)$ from \eqref{eq:mrd} is given by
\begin{equation}
\mathcal{F}(z',z, Z) = \frac{1}{\mathcal{M}_{{\rm{TOT}}}(Z)}\frac{{\rm{d}}\mathcal{N}(z',z, Z)}{{\rm{d}}t(z)}\,,
\end{equation}
where $\mathcal{M}_{{\rm{TOT}}}(Z) = \mathcal{M}_{{\rm{sim}}}(Z)/(f_{\mathrm{bin}}f_{\mathrm{IMF}})$. Here, $\mathcal{M}_{{\rm{sim}}}(Z)$ is the total simulated initial stellar mass in \textsc{sevn},  $f_{\mathrm{bin}} = 0.4$ accounts for the assumed binary fraction \citep{sana2012},  and $f_{\mathrm{IMF}} = 0.255$ accounts for the incomplete sampling of the IMF \citep{iorio2023}. Additionally, ${{\rm{d}}\mathcal{N}(z',z, Z)/{\rm{d}}}t(z)$ denotes the rate of BBH mergers originating from progenitor stars with a metallicity $Z$ at redshift $z'$ that merge at redshift $z$.

The total number of BBH mergers for each model, regardless of whether they are detectable, is reported in Table~\ref{tab:IC}. This is given by
\begin{equation}
\label{eq:n_sources}
    \mathcal{N} =T_{\mathrm{obs}}\int \mathcal{R}(z)p(m_1,m_2|z)\frac{1}{1+z}\frac{\mathrm{d}V_c}{\mathrm{d}z}\mathrm{d}m_1\mathrm{d}m_2
    \mathrm{d}z\,,
\end{equation}
where $\mathcal{R}(z)$ is the merger rate density (\eqref{eq:mrd}), and $p(m_1, m_2|z)$ is the two-dimensional source-frame mass distribution at a given redshift
extracted with \textsc{cosmo$\mathcal{R}$ate} for each astrophysical model (Table~\ref{tab:IC}). The factor $1/(1 + z)$ converts source-frame time into detector-frame time, $\mathrm{d}Vc/\mathrm{d}z$ is the differential comoving volume element, and $T_{\mathrm{obs}}$ is the observing time.

For each of our models, we drew a realization of $\mathcal{N}$ BBHs assuming $T_{\mathrm{obs}} = 10$~yr, obtaining a number of Pop.~III mergers that is orders of magnitude smaller for the fiducial ($\ssim 7\times10^4$), optimistic ($\sim 1\times10^5$), and pessimistic ($\ssim 2\times10^2$) model compared to  Pop.~I-II BBHs ($\ssim 9\times10^5$).

\subsection{Parameter estimation}

\label{sec:pe}

\begin{table}
\caption{Signal parameters and their adopted priors.}    
\label{tab:prior}
\centering
\begin{tabular}{l  c c}
\hline\hline  
Parameter & Units & Prior \\ 
\hline
$m_{1,\mathrm{d}}$ &  [M$_\odot$] & $\mathcal{U}(0, \infty)$ \\
$m_{2, \mathrm{d}}$ &  [M$_\odot$] & $\mathcal{U}(0, m_{1,\mathrm{d}})$ \\
$d_{\mathrm{L}}$ &  [Mpc] & \eqref{eq:priordistance} \\
ra  &[rad] & $\mathcal{U}(0, 2\pi)$\\
$\sin \mathrm{dec}$  &[rad] &  $\mathcal{U}(- 1, 1)$\\
$\cos \iota$ &[rad] &  $\mathcal{U}(-1, 1)$\\
$\phi_c$ & [rad] &  $\mathcal{U}(0, 2\pi)$\\
$\psi$ & [rad] & $\mathcal{U}(0, \pi)$\\
$t_c$ & [s] & $\mathcal{U}(0, \infty)$ \\
\hline
\end{tabular}
\tablefoot{$\mathcal{U}(a,b)$ denotes a uniform distribution between $a$ and $b$. See Sect.~\ref{sec:pe} for details.}
\end{table}

We simulated the parameter-estimation performance of the ET using the FIM formalism \citep{Cutler:1994ys,vallisneri2008,chan2018,grimm2020,Borhanian2021,iacovelli2022b} as implemented in \textsc{GWFish} \citep{dupletsa2023}. In this approach, the likelihood of the data realization $d$ is approximated as a multivariate Gaussian distribution. The posterior is therefore given by 
\begin{equation}
\label{eq:multivar}
    p(\bm{\theta}|d) \propto \pi(\bm{\theta}) \exp{\left(-\frac{1}{2}~(\bm{\theta}-\bm{\overline{\theta})}^\mathrm{T}\, \mathcal{F} \, {(\bm{\theta}-\bm{\overline{\theta}})}\right)}\,,
\end{equation}
where $\bm{\overline{\theta}}$ are the injected parameters. We assumed the prior distributions $\pi(\bm{\theta})$ 
reported in Table~\ref{tab:prior} \citep{Dupletsa2024}. In particular, for the luminosity distance, we chose a prior uniform in comoving volume and source-frame time,
\begin{equation}
\label{eq:priordistance}
    \pi(d_{\mathrm{L}}) \propto \frac{\mathrm{d}V_c}{\mathrm{d}d_{\mathrm{L}}}\frac{1}{1+z} =  \frac{\mathrm{d}V_c}{\mathrm{d}z}\frac{\mathrm{d}z}{\mathrm{d}d_{\mathrm{L}}}\frac{1}{1+z}\,,
\end{equation}
where $d_{\mathrm{L}} \in [0,d_{\mathrm{L}}(z = 1000)]$ Mpc. The FIM is given by
\begin{equation}
\label{eq:fisher}
    \mathcal{F}_{ij} = \left( 
    \frac{\partial h}{\partial \theta _i} 
    \left| 
    \frac{\partial h}{\partial \theta _j} 
\right.\right) \Bigg|_{\bm{\theta}=\bm{\overline{\theta}}} \,,
\end{equation}
where \textsc{GWFish} adopts ${\bm{\theta}} = \{ {m_{1,\mathrm{d}},\,m_{2,\mathrm{d}},\,d_{\mathrm{L}},\,\mathrm{ra},\,\mathrm{dec},\,\iota,\,\phi_c,\,\psi,\,t_c}\}$, where $m_{1,\mathrm{d}}$,~ $m_{2,\mathrm{d}}$, and $d_{\mathrm{L}}$ denote the detector-frame masses and luminosity distance, respectively; ra and dec are sky position coordinates, $\iota$ is the inclination angle of the binary with respect to the line of sight, $\phi_c$ is the phase at coalescence, $\psi$ is the polarization angle, and $t_c$ is the time of coalescence. The waveforms also depend on the spin parameters, which we set to zero. We explain our motivation for this choice and discuss it in Sect.~\ref{sec:spins}. 

We used the waveform approximant \textsc{IMRPhenomHM} \citep{Kalaghatgi2020}. The inner product in \eqref{eq:fisher} is defined as 
\begin{equation}
\label{eq:product}
    ( a | b ) = 4 \Re \int_{f_{\mathrm{low}}}^{f_{\mathrm{high}}} \frac{{a(\bm{\theta},f)b^*(\bm{\theta},f)}}{S_n(f)}\,\mathrm{d}f\,,
\end{equation}
with $f_{\mathrm{low}} = 2$~Hz and $f_{\mathrm{high}} = 2048$~Hz, and $a(\bm{\theta},f)$ and $b(\bm{\theta},f)$ are the Fourier transforms of the time-domain signals $a(\bm{\theta},t)$ and $b(\bm{\theta},t)$. The noise power spectral density $S_n(f)$ was taken to be that of a single triangular ET detector with an arm length of 10 km located in Sardinia (latitude 40$\degree$
$31^\prime$, longitude 9$\degree$ $25^\prime$), which is a possible candidate site \citep{branchesi2023}. We used the HFLF-cryogenic sensitivity curve\footnote{The sensitivity curve can be downloaded at \href{https://apps.et-gw.eu/tds/?content=3&r=14065}{https://apps.et-gw.eu/tds/?content=3\&r=14065}.}, which includes both a high-frequency instrument and a cryogenic low-frequency instrument.

As the signal observed at the detector depends on the redshifted masses and the luminosity distance to the source, we converted the source-frame masses and redshift obtained through 
\textsc{cosmo$\mathcal{R}$ate} in Sect.~\ref{sec:cosmorate} into
detector-frame masses and luminosity distances, 
\begin{align}
\label{eq:detector_frame}
    m_{i,\mathrm{d}} &= m_{i}(1+z)\,,\\
    d_{\mathrm{L}} &= c(1+z) \int_0^z \frac{dz'}{H(z')}\,,
\end{align}
where $i = {1,2}$ denotes the primary and secondary mass, respectively; and the Hubble parameter $H(z)$ is given in \eqref{eq:cosmo}. 

The parameters $m_{1,\mathrm{d}},~ m_{2,\mathrm{d}},$ and $d_{\mathrm{L}}$ were distributed accordingly to their astrophysical populations (Table~\ref{tab:IC}). The other parameters were extracted from their prior distributions (Table~\ref{tab:prior}). Furthermore, we assumed a 100\% duty cycle such that $T_{\mathrm{obs}}$ corresponds to the full data-taking time.

\begin{figure}\
\centering
 \includegraphics[width = 0.95\columnwidth]{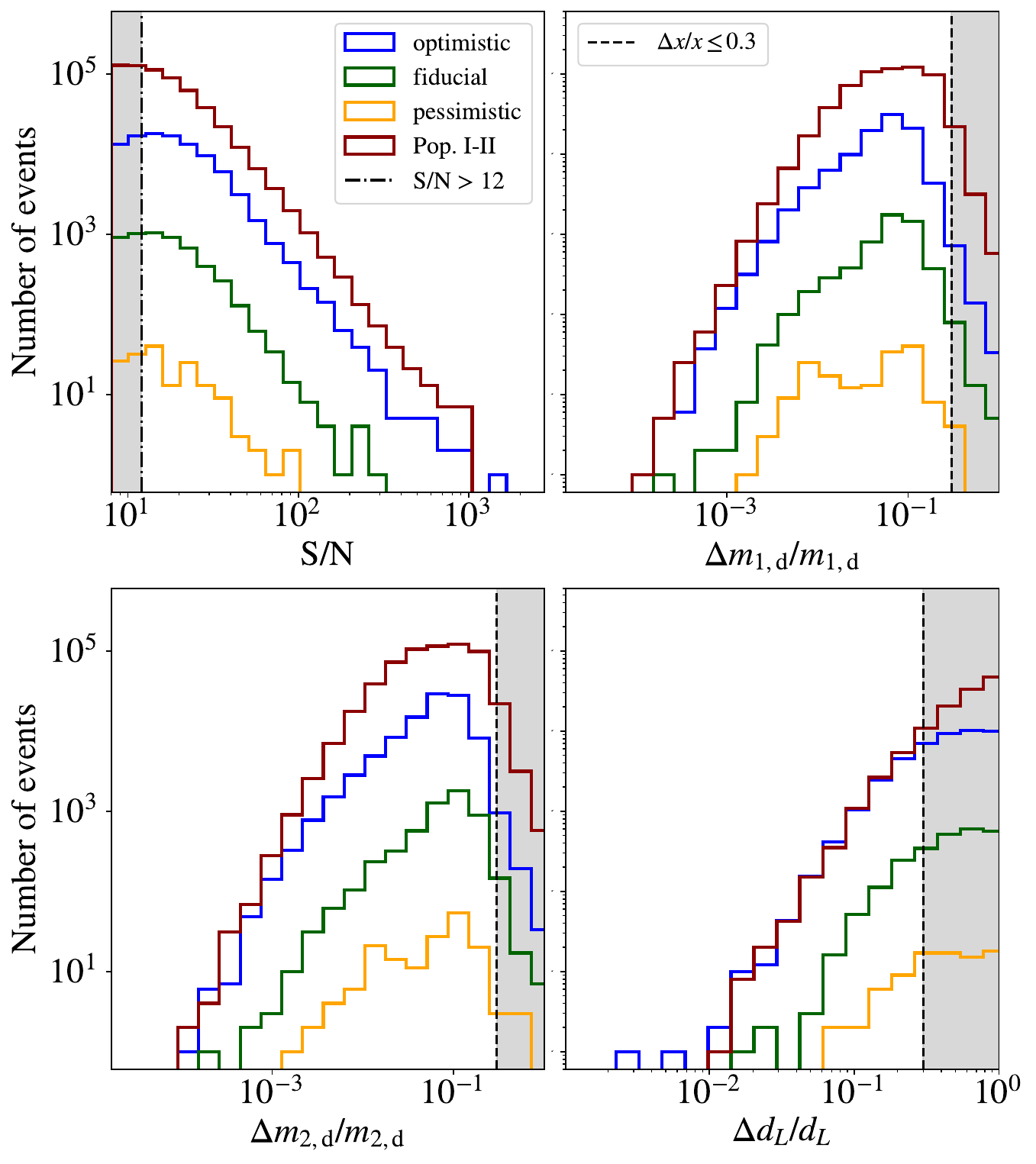}
 \caption{Distribution of S/Ns (top left panel) and relative errors for detector-frame primary masses (top right panel), detector-frame secondary masses (bottom left panel), and luminosity distances (bottom right panel). The blue, green, and orange histograms correspond to optimistic, fiducial, and pessimistic models for Pop.~III BBHs, respectively, and the red histograms refer to BBHs from Pop.~I-II stars (see Sects.~\ref{sec:astro} and \ref{sec:pe} for details). In the top left panel, the dash-dotted black line marks the adopted S/N limit, and the dashed black lines in the other panels represent the adopted threshold for relative errors, where $\Delta x$ are $1\sigma$ errors and $x=\{m_{1,\mathrm{d}},~m_{2,\mathrm{d}}, d_{\mathrm{L}}\}$.}
\label{fig:rel_err}
\end{figure}

In Fig.~\ref{fig:rel_err}, we show the parameter estimation for BBHs originating in Pop.~III and Pop.~I-II stars observed with the ET. 
Notably, a few Pop.~III sources in the optimistic model are measured with extremely high precision. They have $\Delta d_{\mathrm{L}}/d_{\mathrm{L}} \leq 0.01$ because they are at low redshift ($z \lesssim 1$) and have high source-frame masses ($m_1 \gtrsim 30$~M$_\odot$). 

For the remainder of the paper, we select sources with ${\rm S/N} \geq 12$ and relative errors smaller than 0.3 on $m_{1,\mathrm{d}},~m_{2,\mathrm{d}}$ and $d_{\mathrm{L}}$. We further discuss this choice in Sect.~\ref{sec:caveatsonFIM}. As a result, the number of detected and selected Pop.~I-II BBHs is $N_{\mathrm{Pop.\ I-II}} = 11397$ and the number of Pop.~III BBHs is $N_{\mathrm{Pop.\ III}} = 464, 10185, 22$ for the fiducial, optimistic, and pessimistic case, respectively.  Despite these thresholds, our analysis considered sources at low and high redshifts, with maximum redshift values of $z_{\mathrm{max}} \simeq 23, 19,$ and 21 for the fiducial, pessimistic, and optimistic models, respectively. We also created balanced datasets, that is, we increased $T_{\rm{obs}}$ in \eqref{eq:n_sources} to obtain exactly 11397  events for the Pop.~III models as well.

\begin{figure}
\centering
 \includegraphics[width = 0.95\columnwidth]{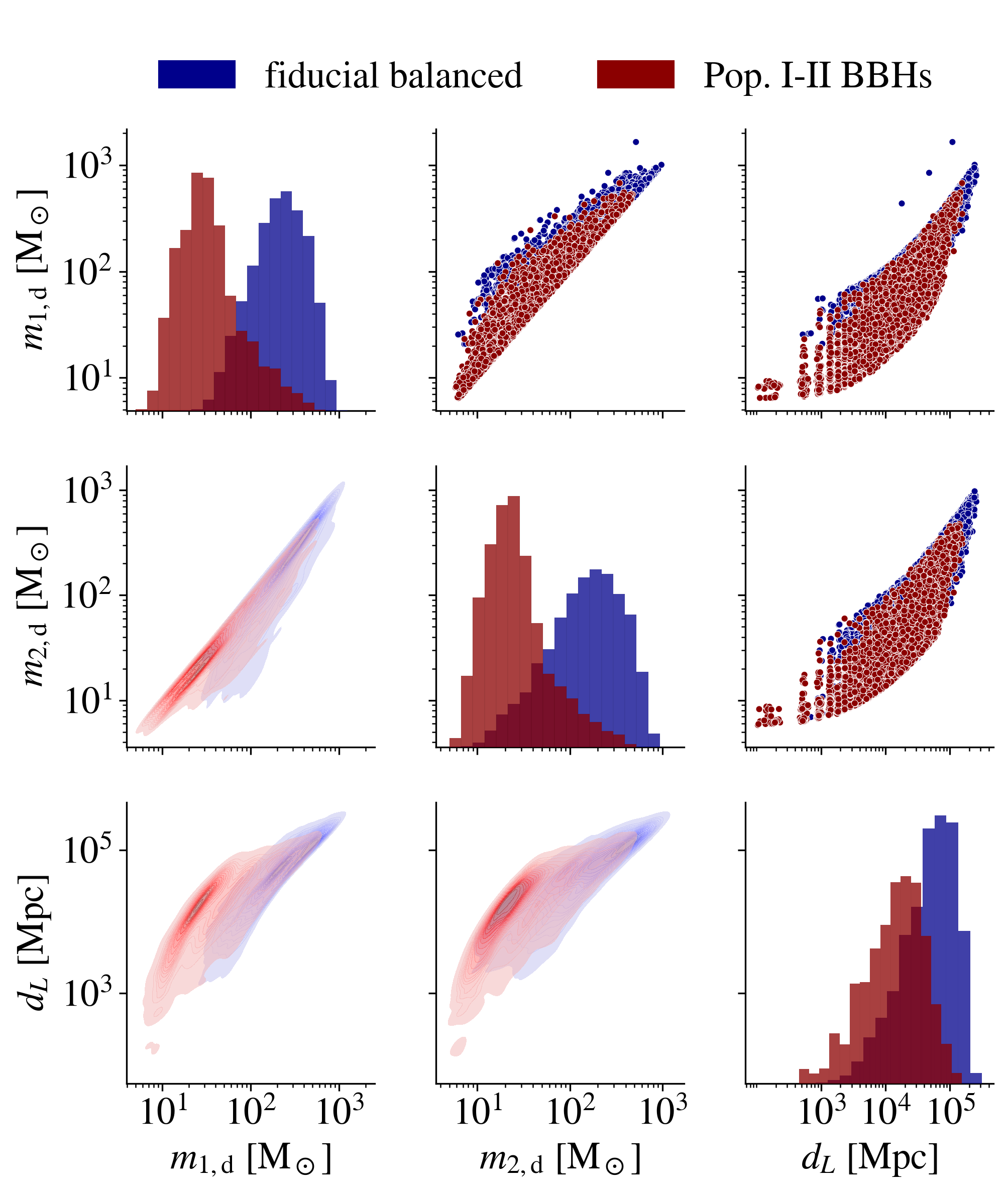}
 \caption{Pair plot illustrating the distribution for the fiducial balanced case. The red and blue distributions indicate mergers of BBHs from Pop.~I-II and Pop.~III stars, respectively. The lower plots show kernel density estimations, the upper plots show the individual samples, and the diagonal plots show the marginalized distributions.}
\label{fig:fid}
\end{figure}

\begin{figure}
\centering
 \includegraphics[width = 0.95\columnwidth]{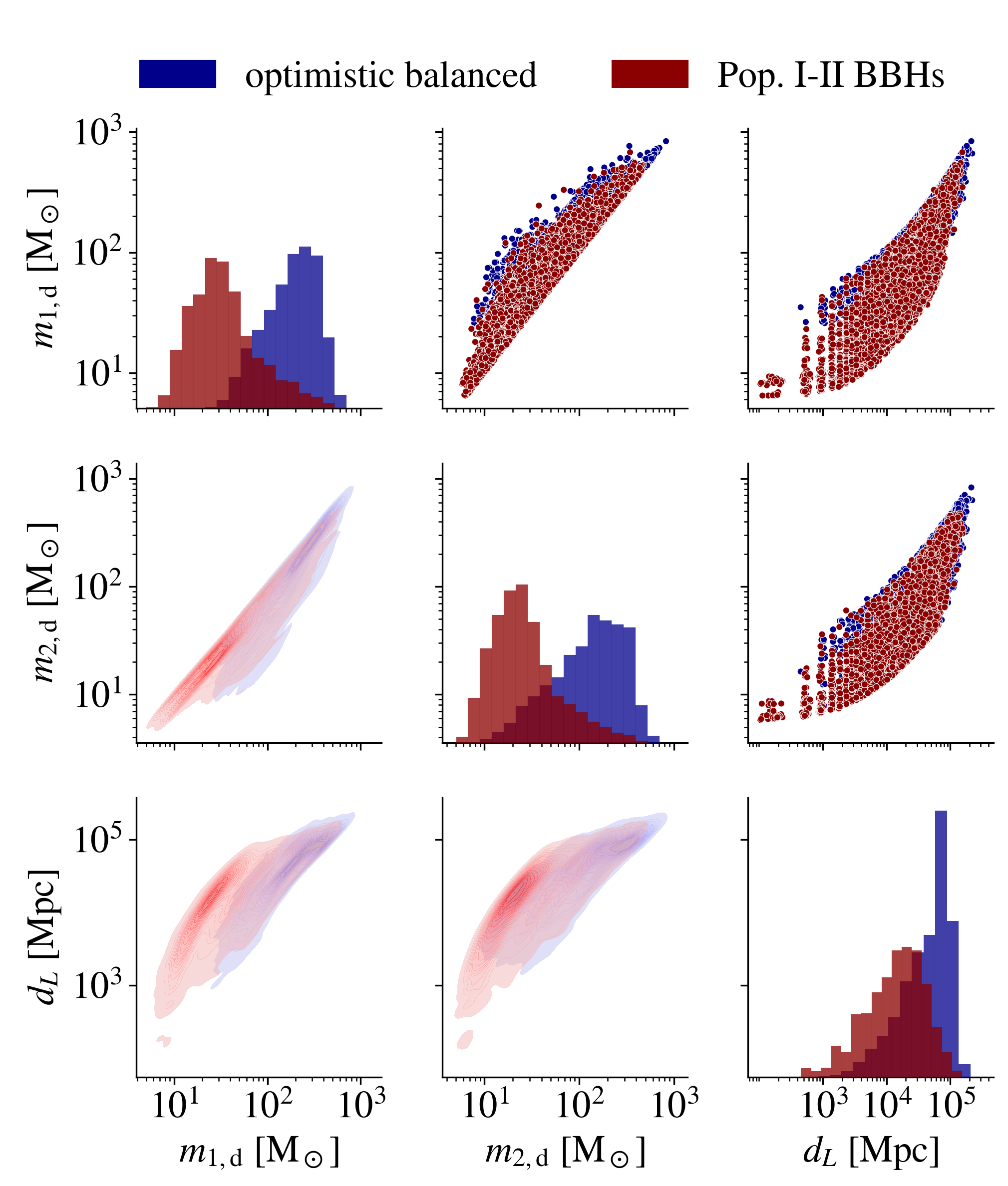}
 \caption{Same as Fig.~\ref{fig:fid}, but for the optimistic balanced model.}
\label{fig:opt}
\end{figure}

\begin{figure}
\centering
 \includegraphics[width = 0.95\columnwidth]{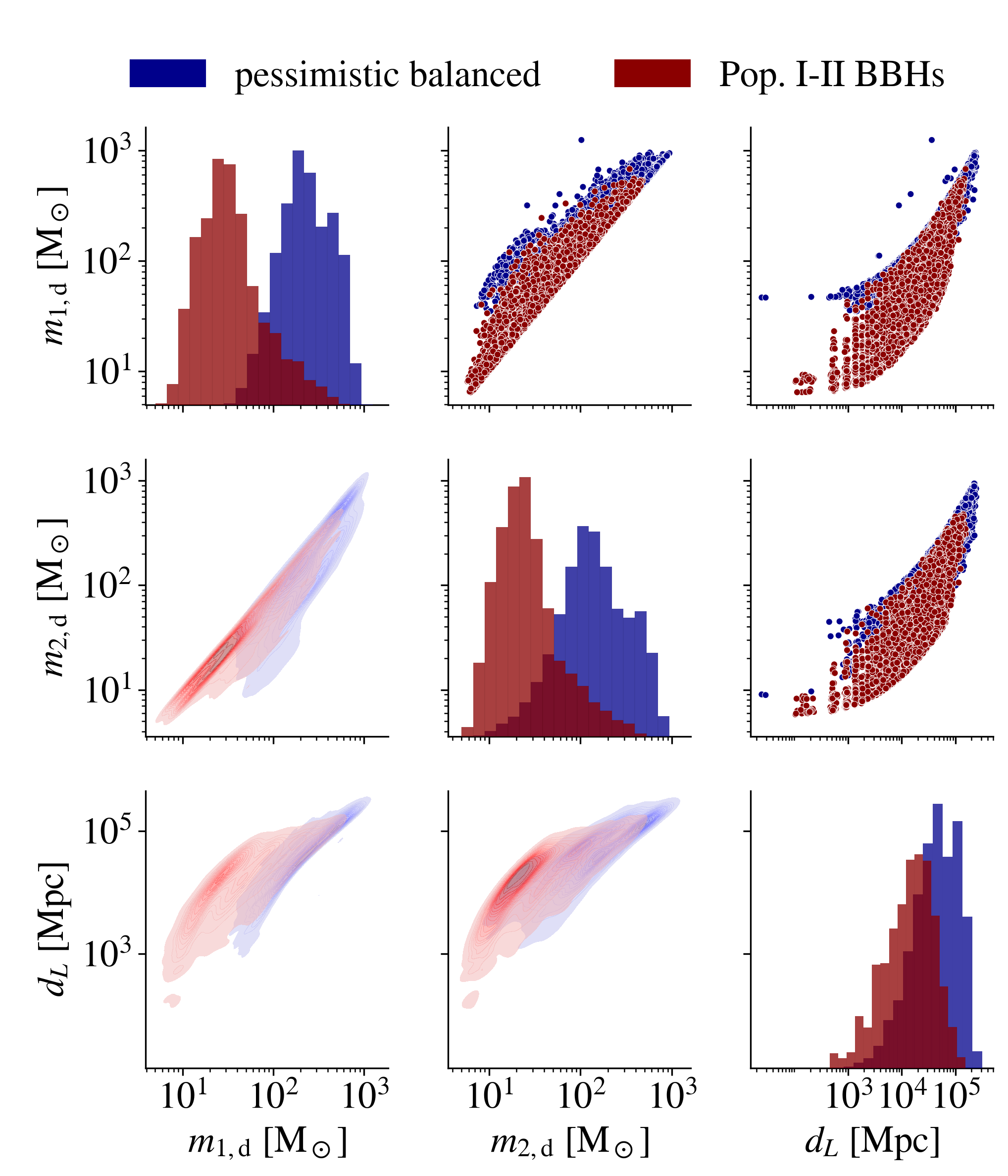}
 \caption{Same as Fig.~\ref{fig:fid}, but for the pessimistic balanced model.}
\label{fig:pes}
\end{figure}

Figs.~\ref{fig:fid}, \ref{fig:opt}, and \ref{fig:pes} illustrate the parameter distributions of Pop.~III BBHs in the fiducial, optimistic, and pessimistic balanced models compared to those of Pop.~I-II. The one-dimensional histograms along the diagonal axes of the pair plots show distinct peaks in the distributions of all the parameters. For instance, in the detector frame, the primary mass peaks at $20-30$~M$_\odot$ for Pop.~I-II BBHs, whereas for Pop.~III objects, the maximum is at $200-300$~M$_\odot$. On average, Pop.~III BBHs exhibit detector-frame masses that are approximately 100 times higher than those of Pop.~I-II. The peak of the source-frame primary mass distribution for BBHs generated from Pop.~III stars falls within the range of $m_1 \sim 30 - 40$~M$_\odot$, and the peak of the merger rate density occurs at $z \sim 8 - 16$  \citep{santoliquido2023}. In the detector frame, the primary masses of Pop.~III BBHs are therefore  $m_{1,\mathrm{d}} \sim 240 - 640$~M$_\odot$. In contrast, Pop.~I-II BBHs exhibit a peak in the source-frame primary mass distribution at $10-15$ M$_\odot$ \citep{broek2022,vanson2022,santoliquido2023} , and the merger rate density peaks at $z \sim 2-3$ \citep{santoliquido2021}. 
The secondary mass distribution of Pop.~I-II objects peaks at $\ssim 20-30$~M$_\odot$, in contrast with Pop.~III, for which it peaks at $\ssim100-200$~M$_\odot$. The luminosity distance peaks at $\ssim 2\times 10^4$~Mpc for Pop.~I-II BBHs, and for Pop.~III sources, it peaks at $\ssim 10^5$~Mpc.

Nevertheless, the two populations overlap: The tails of the distributions of Pop.~I-II $m_{1,\mathrm{d}}$, $m_{2,\mathrm{d}}$, and $d_{\mathrm{L}}$ intersect with the maxima of Pop.~III. This suggests that it is not trivial to efficiently distinguish between these populations.

\subsection{Classification of a single detection}
\label{sec:post_samples}

The probability that the detected event $j$ belongs to the Pop.~III class (hereafter, $k$ for brevity) is given by
\begin{equation}
\label{eq:theequation}
    p(j\in k | d_j, \{\beta\}) = 
    \int \mathrm{d}\mathbf{x} p ( j\in k | \mathbf{x}, d_j,\{\beta\})p (\mathbf{x}| d_j, \{\beta\})\,,
\end{equation}
where $d_j$ is the data stream of the event $j$, $\{\beta\} = \{\beta_{\mathrm{I-II}},~ \beta_{\mathrm{III}}\}$ are the fractions of Pop.~I-II and Pop.~III BBH mergers, respectively, and $\mathbf{x} $ is a subset of the waveform parameters (see Sect.~\ref{sec:classifier}). The term $p ( i\in k | \mathbf{x}, d_i,\{\beta\})$ is the probability that the event $j$ belongs to the Pop.~III class given the waveform parameters $\mathbf{x}$, the data stream $d_j$, and the mixing fractions. We evaluated it by using a ML algorithm trained on detectable waveform parameters (see Sect.~\ref{sec:classifier}).

We approximated the integral in \eqref{eq:theequation} with a Monte Carlo summation. We made the simplifying assumption that the posterior samples of the single event $j$ do not depend on the mixing fractions ${\beta}$, that is, they were obtained using uninformative priors. We further discuss this choice in Sect.~\ref{sec:hyperunc}. Therefore, to solve \eqref{eq:theequation}, we  drew $N_s$ posteriors samples from $p(\mathbf{x}|d_j)$ that are given as in \eqref{eq:multivar} for each detection $j$,
\begin{equation}
    p(j\in k | d_j, \{\beta\}) \approx \left\langle p ( j\in k | \mathbf{x}_i, d_j,\{\beta\}) \right\rangle_{\mathbf{x}_i \sim p(\mathbf{x}|d_j)}\,,
\end{equation}
where $i$ spans the $N_s$ posterior samples. The presence of noise typically leads to a deviation of the likelihood maximum (\eqref{eq:multivar}) from the injected values in \textsc{GWFish} \citep{rodriguez2013, Gupta2022}. To simulate this, the $N_s$ posterior samples were centered around a new maximum likelihood that was randomly sampled from \eqref{eq:multivar} for each detected source (for further details, see Section 2.1 in \citealt{iacovelli2022b}). 

\subsection{Classification probability of a single event}
\label{sec:classifier}

\begin{table}
\caption{Hyperparameters adopted in \textsc{XGBoost} training after a randomized search.}        
\label{tab:hyper}      
\centering                          
\begin{tabular}{l l l l l}        
\hline\hline     
Hyperparameter & fid. {bal.} & opt. {bal.} & pes. {bal.} & range \\
\hline
 {\texttt{max\_depth}} &    13 & 13 & 12 & [2, 15]\\
 {\texttt{learning\_rate}} & 0.034  & 0.034 & 0.016 & [0, 0.1]\\
 {\texttt{gamma}} & 0.36 & 0.36 & 0.25 &[0, 3]\\ 
  \hline                                          
\end{tabular}
\tablefoot{The hyperparameters are distributed log-uniformly in the provided ranges. The column names stand for fiducial (fid.), optimistic (opt.), pessimistic (pes.), and balanced (bal.)}
\end{table}

The determination of the probability that the event $j$ belongs to the Pop.~III class given its fixed waveform and population parameters [$p(j \in k| \mathbf{x}, d_j, {\beta})$ in \eqref{eq:theequation}] can be seen as a classification problem that is easily addressed with a ML algorithm. The classification we performed, despite it can be done on every waveform parameter in principle (see Sect.~\ref{sec:pe}), is based only on the detector-frame primary and secondary mass and on the luminosity distance, that is, $\mathbf{x} = \{m_{1,\mathrm{d}},m_{2,\mathrm{d}}, d_{\mathrm{L}}\}$. These parameters are those linked to astrophysical processes, and the remaining parameters were assigned randomly in our simulations.

We trained three different classifiers for which we kept the BBHs from Pop.~I-II stars fixed and varied those from Pop.~III stars (fiducial, optimistic, and pessimistic). We trained the classifiers on balanced datasets because classification performances are maximized when the number of elements in the two classes does not vary by some orders of magnitude \citep{HAIXIANG2017220}. Following customary practice, we reserved 70\% of the sources of the balanced datasets for training and validation and used the remaining 30\% to test the performance of the balanced classifiers.

To perform the classification, we adopted the ML algorithm \textsc{XGBoost} (eXtreme Gradient Boosting; \citealt{xgboost}). The key idea behind \textsc{XGBoost} is to sequentially add shallow decision trees to an ensemble of learners, each of which corrects its predecessor. At every iteration, this algorithm fits the new decision tree to the residual errors from the previous iterations.

The performance of \textsc{XGBoost} relies on a set of user-adjustable hyperparameters aimed at enhancing the classifier performance on a specific dataset. To determine the optimal hyperparameters, we employed the \textsc{RandomizedSearchCV} method from \textsc{sklearn} \citep{scikit-learn}, which randomly selects a fixed number of hyperparameter combinations from a predefined search space. This random sampling makes the process computationally efficient and well suited for extensive search spaces \citep{Bergstra2012,lones2024avoid}. We looped over three hyperparameters. Table~\ref{tab:hyper} outlines the ranges we tested and the identified optimal values. In particular, \texttt{max\_depth} is the maximum depth of a tree in {\sc{XGBoost}}. Increasing \texttt{max\_depth} allows the model to learn more complex relations in the data, but can also lead to overfitting, where the classifier learns the training data excessively, but struggles to generalize to new so-far unseen data. \texttt{learning\_rate} determines the size of the steps taken during training. A lower \texttt{learning\_rate} makes the model more robust, but requires more iterations to converge. \texttt{gamma} also controls overfitting by penalizing the complexity of the trees: higher values of \texttt{gamma} lead to fewer splits. 

To prevent overfitting, we also employed a $k$-fold cross-validation. We divided the training dataset into $k = 5$ subsets. The model undergoes training $k$ times, with each iteration using $k-1$ subsets for training and the remaining set for validation. Implemented for additional robustness, we employed the so-called nested cross-validation, which introduces an inner level of cross-validation with $k = 2$ during the hyperparameter tuning process.

When trained on balanced datasets, \textsc{XGBoost} provides $p(j\in k| \mathbf{x}, d_j)$, a probability independent of the mixing fraction. To evaluate the performances of the balanced classifiers, we considered a fixed threshold, that is, whenever $p(j\in k| \mathbf{x}, d_j) > 0.5$, the $j$ event is predicted  to be a Pop.~III BBH. 

The results are presented using confusion matrices that consist of four entries: true positive (hereafter TP, correctly predicted Pop.~III BBHs), true negative (hereafter TN, correctly predicted Pop.~I-II BBHs), false positive (hereafter FP, incorrectly predicted Pop.~III), and false negative (hereafter FN, incorrectly predicted Pop.~I-II). We evaluated the performance of our classifiers using  precision, recall, and F1 score. Precision is defined as 
\begin{equation}
    \label{eq:precision}
    \mathrm{Precision} = \frac{\mathrm{TP}}{\mathrm{TP}+\mathrm{FP}}\,,
\end{equation}
where a high value indicates that a positive prediction from the classifier is likely to be correct. When TP = 0, we set Precision = 0. Recall (or sensitivity) is defined as
\begin{equation}
\label{eq:recall}
    \mathrm{Recall} = \frac{\mathrm{TP}}{\mathrm{TP}+\mathrm{FN}}\,
\end{equation}
and measures the ability of the classifier to capture all the positive instances (i.e., Pop.~III), with a high value indicating that the model is effective at identifying most of the positive instances. The F1 score 
\begin{equation}
\label{eq:f1}
    \mathrm{F1} = 2~ \frac{\mathrm{Precision}\times\mathrm{Recall}}{\mathrm{Precision}+\mathrm{Recall}}\,
\end{equation}
is the harmonic mean of the precision and recall, providing a balance between the two.

Training a ML classifier on artificially balanced dataset introduces a bias favoring the minority class \citep{weiss}. Among the suggested mitigation strategies \citep{tian2020posterior}, we opt for the following Bayesian approach, which relies on simulation-informed priors \citep{chan2019application,berbel2023}.
The probability we wished to evaluate is $p(j\in k| \mathbf{x}, d_j, {\beta})$, which enters \eqref{eq:theequation}. We started from the Bayes theorem written in terms of odds,
\begin{equation}
\label{eq:odds}
    \frac{p ( i\in k | \mathbf{x}, d_i,\{\beta\})}{p ( i \notin k | \mathbf{x}, d_i,\{\beta\})} = \frac{\pi(i \in k| \{\beta\})}{\pi(i \notin k| \{\beta\})} \frac{p (\mathbf{x}, d_i|i \in k ,\{\beta\})}{p (\mathbf{x}, d_i|i\notin k ,\{\beta\})}\,.
\end{equation}
The likelihood of the single event $j$, denoted as $p (\mathbf{x}, d_i|i \in k ,\{\beta\})$, does not depend on the fractions $\{\beta\}$, which can only vary its normalization.
We can then write $p (\mathbf{x}, d_i|i \in k ,\{\beta\}) = p (\mathbf{x}, d_i| i \in k )$ and apply the Bayes theorem once more to obtain
\begin{equation}
\label{eq:bayes_balanced}
p(i \in k | \mathbf{x}, d_i) = \frac{p(\mathbf{x}, d_i | i \in k) \pi (i \in k)}{p(\mathbf{x}, d_i)} 
= \frac{p(\mathbf{x}, d_i | i \in k)}{\sum _k p(\mathbf{x}, d_i | i \in k)}\,,
\end{equation}
where the second equality follows from the fact that  $\pi(i\in k) = 1/2$. We substitute \eqref{eq:bayes_balanced} twice in \eqref{eq:odds}, making use of the fact that the normalization $\sum_k p(\mathbf{x}, d_i | i \in k)$ is independent of $k$ and therefore cancels out. The relevant prior probabilities are $\pi(i \notin k | \{\beta \}) = \beta_{\mathrm{I-II}}$ and $\pi(i \in k | \{\beta\}) = \beta_{\mathrm{III}}$. 
\eqref{eq:odds} then becomes
\begin{equation}
\label{eq:rebalancing}
    p ( i\in k | \mathbf{x}, d_i,\{\beta\}) = \frac{p( i \in k | \mathbf{x}, d_i)\beta_{\mathrm{III}}}{p( i \in k | \mathbf{x}, d_i)(\beta_{\mathrm{III}}-\beta_{\mathrm{I-II}}) + \beta_{\mathrm{I-II}}}\,,
\end{equation}
where $\beta_{\mathrm{III}} = N_{\mathrm{Pop.~III}}/N_{\mathrm{tot}}$ and $\beta_{\mathrm{I-II}} = N_{\mathrm{Pop.~I-II}}/N_{\mathrm{tot}}$ with $N_{\mathrm{tot}} = N_{\mathrm{Pop.~III}} + N_{\mathrm{Pop.~I-II}} $ (see Sect.~\ref{sec:pe})

\section{Results}
\label{sec:results}

\subsection{Classification performances on balanced datasets}
\label{sec:res_training}

\begin{figure}
\centering
\includegraphics[width = 0.4\textwidth]{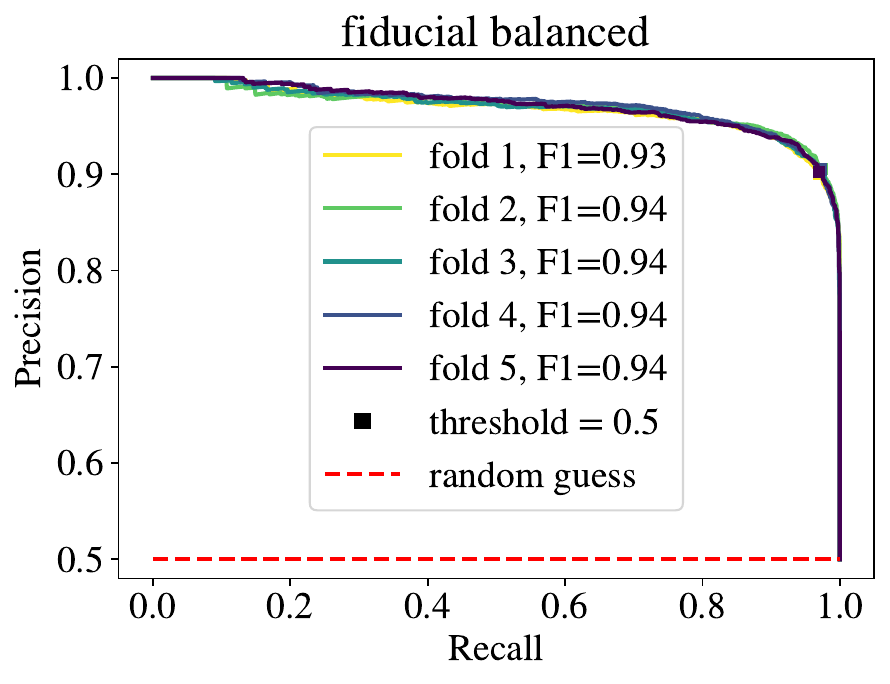}\hfill
\includegraphics[width = 0.4\textwidth]{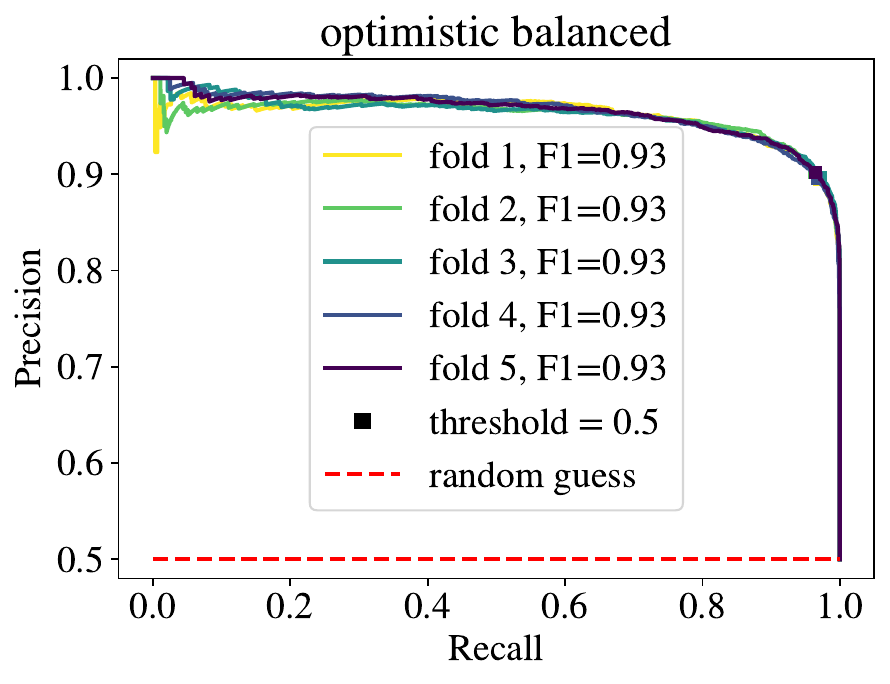}\hfill
\includegraphics[width = 0.4\textwidth]{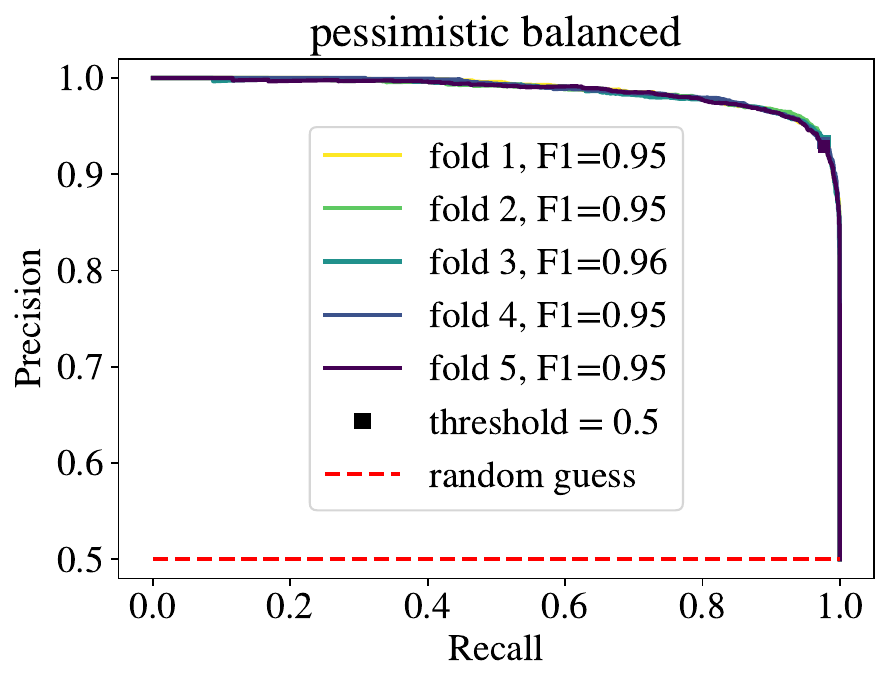}
\caption{Precision-recall curves evaluated on the test sets for the fiducial balanced model (top panel), optimistic balanced (middle panel), and  pessimistic balanced (bottom panel). The colors indicate the results we obtained for different folds. F1 is reported as a scoring parameter for each fold and has been evaluated by applying a threshold equal to 0.5 on $p(j \in k |\mathbf{x}, d_j)$ (see Sect.~\ref{sec:classifier} for details). The dashed horizontal red line represents the precision-recall curve of a classifier making random guesses. }
\label{fig:cv}
\end{figure}

Fig.~\ref{fig:cv} shows that the precision-recall curves evaluated on the test sets and with the best hyperparameters listed in Table~\ref{tab:hyper} remain  consistent across the five folds. Moreover, the classifiers exhibit almost the same performances for different data splits, indicating strong generalization capabilities to unseen data.

\begin{table}
\caption{Scoring parameters evaluated on the balanced test sets for the three classifiers of Pop.~III BBHs.}
    \label{tab:performance}
    \centering
    \begin{tabular}{l c c c}
    \hline
    \hline
        Model & Precision & Recall & F1 score\\
        \hline
         Fiducial balanced &0.90 & 0.97 & 0.94 \\
        Optimistic balanced & 0.90 &  0.96 & 0.93\\
         Pessimistic balanced &0.92 &0.98 & 0.95 \\
        \hline
    \end{tabular}
    \tablefoot{See Sect.~\ref{sec:classifier} for details}
\end{table}


\begin{figure*}
\centering
\includegraphics[width = 0.3\textwidth]{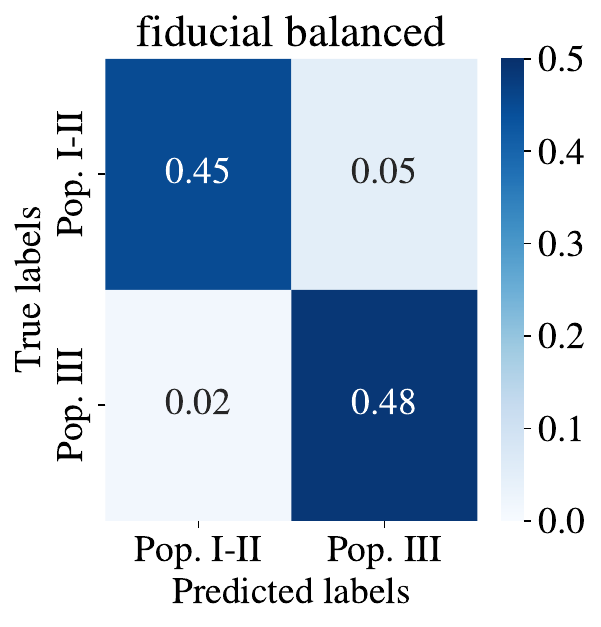}\hfill
\includegraphics[width = 0.3\textwidth]{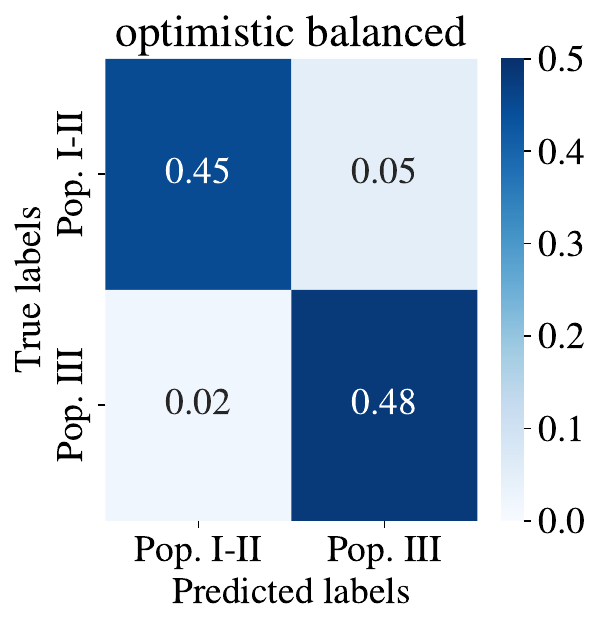}\hfill
\includegraphics[width = 0.3\textwidth]{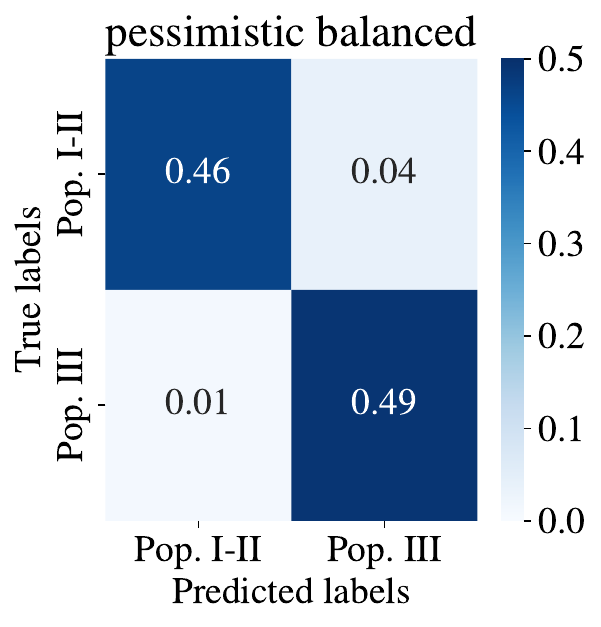}
\caption{Confusion matrix evaluated on the test sets for the fiducial balanced model (left panel), optimistic balanced model (middle panel), and pessimistic balanced model (right panel). The entries are color-coded and normalized to one.}
\label{fig:confs}
\end{figure*}


The confusion matrices evaluated on the test sets are presented in Fig.~\ref{fig:confs} and indicate good  classification performances across all three models. The majority of instances are correctly classified, as shown with high fractions $(\geq 0.45)$ on the diagonal of each matrix. These robust performances are further corroborated by the scoring parameters evaluated on the test sets presented in Table~\ref{tab:performance}. In particular, the achieved precision evaluated on the test sets consistently exceeds 0.90 for each classifier.

The classifiers exhibit a higher rate of confusing Pop.~I-II BBHs as Pop.~III than of the reverse (top right corners of the confusion matrices). This is because the tails of the parameter distribution ($m_{1,\mathrm{d}}$, $m_{2,\mathrm{d}}$, and $d_{\mathrm{L}}$) of Pop.~I-II BBHs extend in the ranges occupied by Pop.~III BBHs (see Figs.~\ref{fig:fid}, \ref{fig:opt}, and \ref{fig:pes}). This fact might imply worse classifier performances in the presence of BBHs that were generated through alternative formation channels, as discussed in more detail in Sect.~\ref{sec:hyperunc}.

\subsection{Classification of single detections}

When we take the expected merger rate ratio between detected Pop.~III and Pop.~I-II BBHs into account, we count that $\ssim 4\%$, $\ssim 47$\%, and $\ssim 0.2\%$ of the total number of BBH mergers are generated from the fiducial, optimistic, and pessimistic model, respectively (Sect.~\ref{sec:pe}).  We used the probability defined in \eqref{eq:theequation}, which provides a confidence in determining whether a source belongs to the Pop.~I-II or Pop.~III class. We can thus introduce a threshold on $p(j\in k | d_j, \{\beta\})$: A higher threshold results in fewer Pop.~III BBHs being identified, but with higher certainty. We evaluated \eqref{eq:theequation} only for those sources that were in the test sets.

Table~\ref{tab:fid} presents the scores as function of this threshold. Our results reveal that we can effectively distinguish BBHs from Pop.~III stars with high precision. For instance, Table~\ref{tab:fid} illustrates that with a threshold equal to 0.7, we can identify $\gtrsim 10\%$ of Pop.~III BBHs in the fiducial model with a precision $>90\%$.

\begin{figure*}
\centering
 \includegraphics[width = 0.85\textwidth]{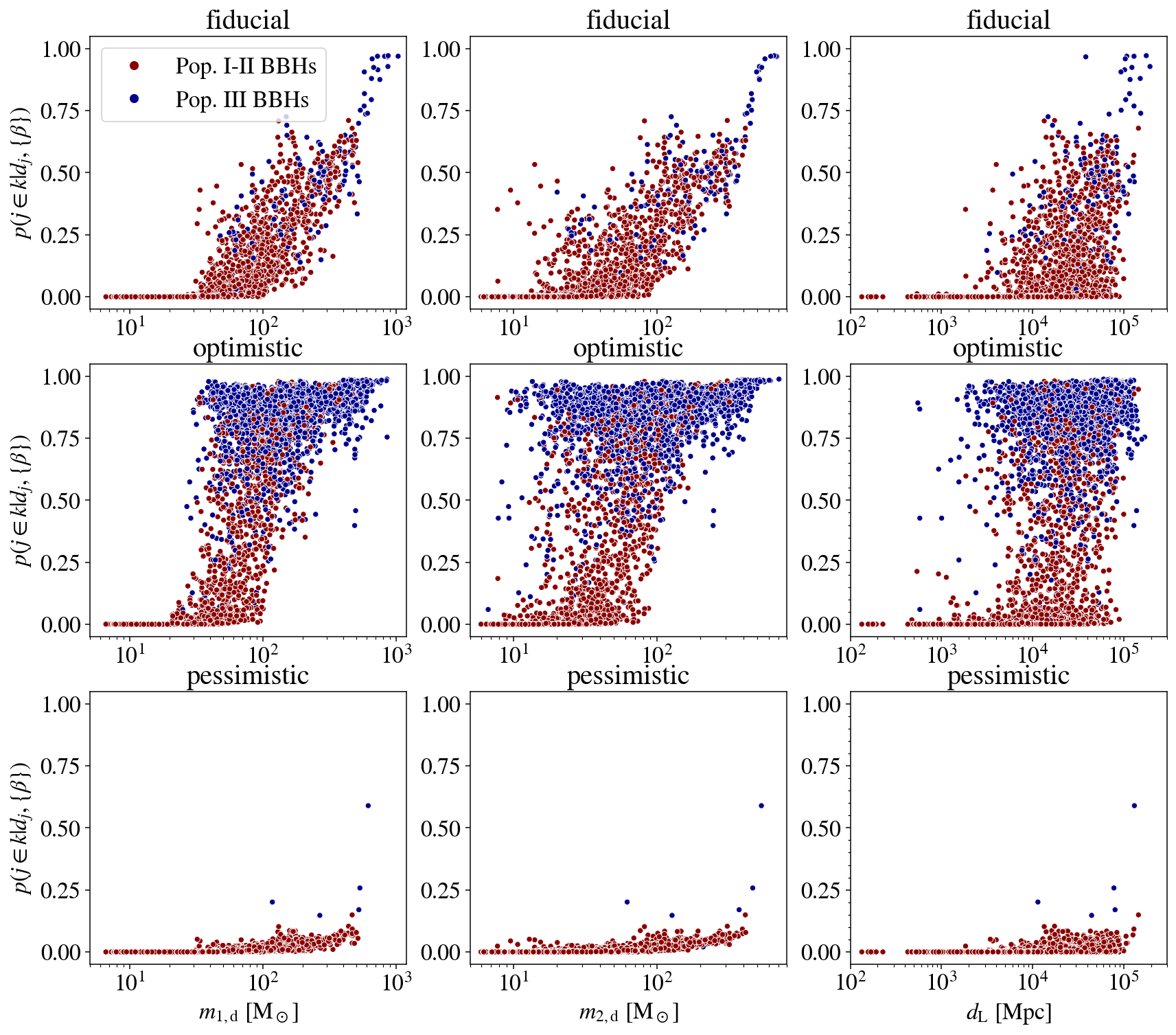}
 \caption{Probability ($p(j\in k | d_j, \{\beta\})$  in \eqref{eq:theequation}) of each mock observation to belong to the Pop.~III class in the left to right columns as a function of primary mass ($m_{1,\mathrm{d}}$), secondary mass ($m_{2,\mathrm{d}}$), and luminosity distance ($d_{\mathrm{L}}$). Each dot corresponds to the median of the posterior samples. The top, middle, and bottom rows correspond to the fiducial, optimistic, and pessimistic scenarios, respectively. The red (blue) dots indicate observations known to belong to Pop.~I-II (Pop.~III). We refer to Sect.~\ref{sec:post_samples} for details.}
\label{fig:prob}
\end{figure*}

As shown in Fig.~\ref{fig:prob}, sources that are confidently identified as Pop.~III BBHs, that is, sources with high $p(j\in k | d_j, \{\beta\})$, typically exhibit a high median detector-frame primary mass ($m_{1,\mathrm{d}} \gtrsim 500$~M$_\odot$) and secondary mass ($m_{2,\mathrm{d}} \gtrsim 400$~M$_\odot$) as well as high median luminosity distance ($d_{\mathrm{L}} \gtrsim 1\times10^5$~Mpc).

\begin{table}
\caption{Scoring of the fiducial (top), optimistic, (middle) and pessimistic (bottom) models for different thresholds on $p(j\in k | d_j, \{\beta\})$, as reported in the first column. } 
    
\label{tab:fid}
\centering
\textbf{Fiducial} \\[5pt]
\resizebox{\columnwidth}{!}{\begin{tabular}{c c c c c c c }
\hline\hline  
Thr. & \%TP & \%TN & \%FP & \%FN & Precision & Recall \\
\hline
0.1 & 98 & 83 & 17 & 2 & 0.19 & 0.98 \\ 
0.2 & 92 & 89 & 12 & 8 & 0.25 & 0.92 \\ 
0.5 & 38 & 98 & 3 & 62 & 0.43 & 0.38 \\ 
0.7 & 14 & 100 & 1 & 87 & 0.90 & 0.14 \\ 
0.9 & 7 & 100 & 0 & 93 & 1.00 & 0.07 \\ 
\hline
\end{tabular}}
$\,$\\[5pt]\textbf{Optimistic} \\[5pt]
\centering
\resizebox{\columnwidth}{!}{\begin{tabular}{c c c c c c c }
\hline\hline  
Thr. & \%TP & \%TN & \%FP & \%FN & Precision & Recall \\
\hline
0.1 & 100 & 74 & 26 & 1 & 0.78 & 1.00 \\ 
0.2 & 100 & 77 & 23 & 1 & 0.79 & 1.00 \\ 
0.3 & 99 & 79 & 21 & 1 & 0.81 & 0.99 \\ 
0.5 & 97 & 83 & 17 & 3 & 0.84 & 0.97 \\ 
0.7 & 91 & 88 & 12 & 9 & 0.87 & 0.91 \\ 
0.9 & 51 & 95 & 5 & 49 & 0.91 & 0.51 \\ 
\hline
\end{tabular}}
$\,$\\[5pt]\textbf{Pessimistic} \\[5pt]
\centering
\resizebox{\columnwidth}{!}{\begin{tabular}{c c c c c c c }
\hline\hline  
Thr. & \%TP & \%TN & \%FP & \%FN & Precision & Recall \\
\hline
0.1 & 83 & 100 & 0 & 17 & 0.62 & 0.83 \\ 
0.2 & 50 & 100 & 0 & 50 & 1.00 & 0.50 \\ 
0.3 & 17 & 100 & 0 & 83 & 1.00 & 0.17 \\ 
0.5 & 17 & 100 & 0 & 83 & 1.00 & 0.17 \\ 
0.7 & 0 & 100 & 0 & 100 & 0 & 0 \\ 
0.9 & 0 & 100 & 0 & 100 & 0 & 0 \\ 
\hline
\end{tabular}}
\tablefoot{The scoring metrics are reported in Sect.~\ref{sec:classifier}}
\end{table}

\section{Discussion}
\label{sec:discussion}

\subsection{Interpretability and manual classifier}
\label{sec:manual}

\begin{figure*}
\centering
\includegraphics[width = 0.3\textwidth]{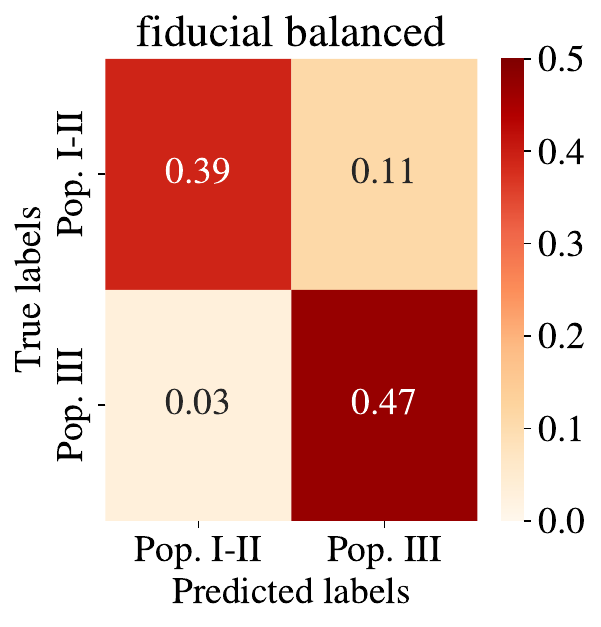}\hfill
\includegraphics[width = 0.3\textwidth]{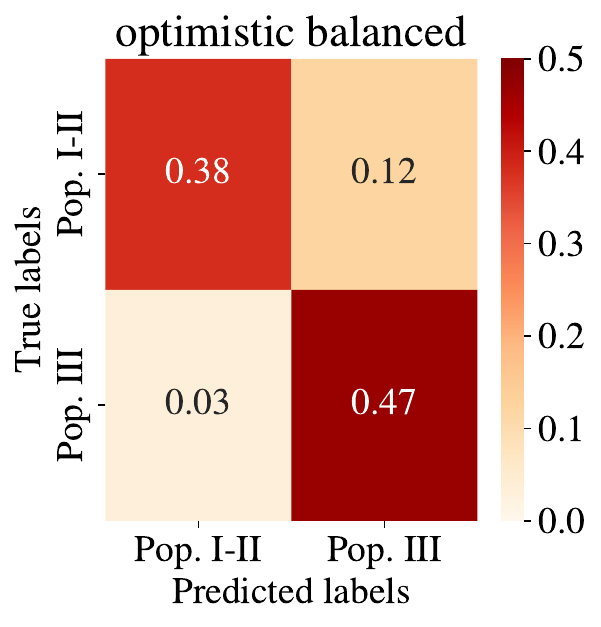}\hfill
\includegraphics[width = 0.3\textwidth]{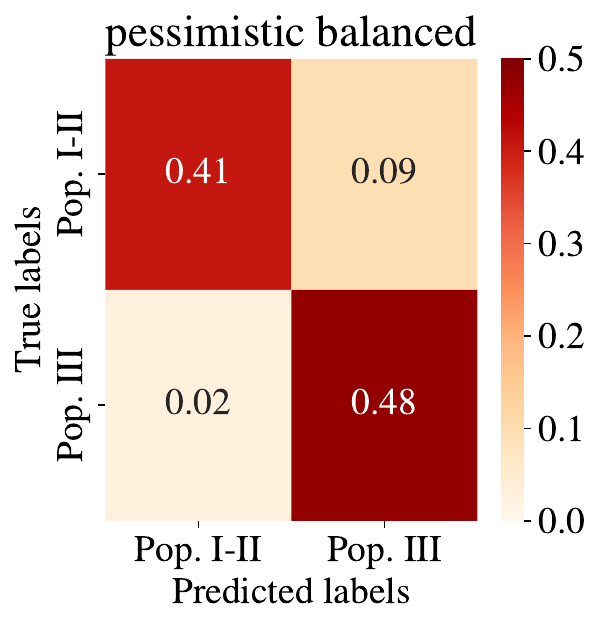}
\caption{Confusion matrix for the balanced fiducial model (left panel), balanced optimistic  (middle panel), and balanced pessimistic (right panel) adopting manual classification (see Sect.~\ref{sec:manual} for details). The entries are color-coded and normalized to one.}
\label{fig:manual}
\end{figure*}

One of the well-known limitations of ML algorithms is their lack of interpretability; that is, it is not always clear why or on which basis an algorithm predicts a specific outcome. Various methods exist to address this issue with various sophistication \citep{angelino2018learning,molnar2021general,Angelov2021ExplainableAI,9050829}. However, given the simplicity of the problem we aim to solve here, that is, to classify BBHs based on only three features, we can opt for a straightforward but highly interpretable approach: a manual classifier.

We devised a simple algorithm that systematically explores a fine grid of values within the feature space. The objective of this manual classifier is to identify thresholds for the primary mass, secondary mass, and luminosity distance that yield the highest precision [\eqref{eq:precision}]. We used the same balanced training and testing dataset as described in Sect.~\ref{sec:classifier}.

The results of this approach were straightforward:
We found optimal scores for the manual classifier that were exclusively determined by the threshold on the detector-frame primary mass, and the secondary mass and luminosity distance were just required to be above zero.

For the fiducial, optimistic, and pessimistic models, the thresholds we found are $m\substack{\mathrm{thr}\\{1,\mathrm{d}}} \simeq 62$~M$_\odot$, $m\substack{\mathrm{thr}\\{1,\mathrm{d}}} \simeq 57$~M$_\odot$, and $m\substack{\mathrm{thr}\\{1,\mathrm{d}}} \simeq 72$~M$_\odot$, respectively. This cut-based manual classifier assigns any source with $m_{1,\mathrm{d}} > m\substack{\mathrm{thr}\\{1,\mathrm{d}}}$ to the class of Pop.~III BBHs with $p(j\in k| \mathbf{x}, d_j) = 1$. 

The performances of this manual classifier are comparable to those of \textsc{XGBoost}, with the corresponding confusion matrices reported in Fig.~\ref{fig:manual}. The F1 scores [\eqref{eq:f1}] evaluated on the balanced test sets are 0.87, 0.86, and 0.89 for the fiducial, optimistic, and pessimistic models, respectively. 

\begin{figure}
    \centering
    \includegraphics[width=0.9\columnwidth]{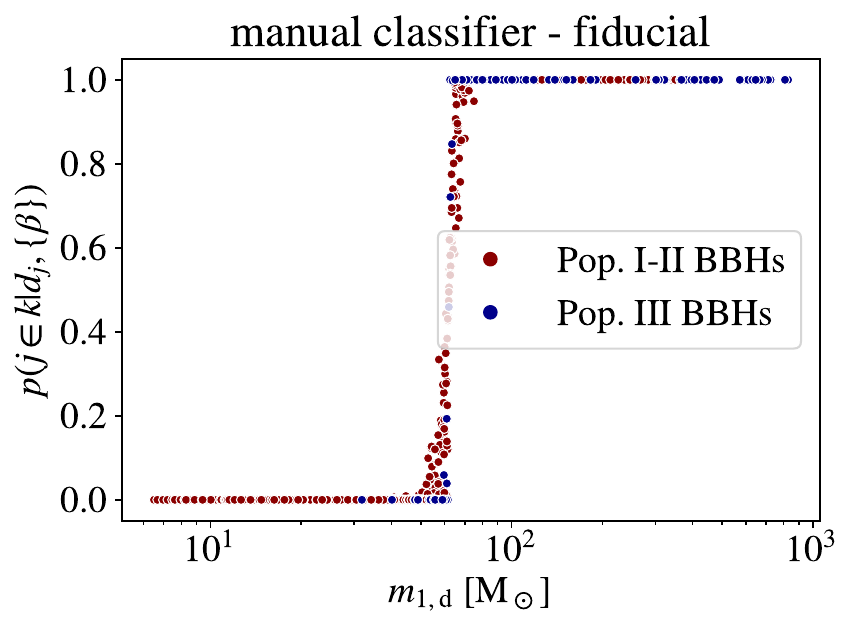}
    \caption{Probability ($p(j\in k | d_j, \{\beta\})$ in \eqref{eq:theequation}) of each mock observation of the fiducial model to belong to the Pop.~III class as a function of primary mass ($m_{1,\mathrm{d}}$),  evaluated with the manual classifier (see Sect.~\ref{sec:manual}). Each dot corresponds to the median of the posterior samples. The red (blue) dots indicate observations known to belong to Pop.~I-II (Pop.~III).}
    \label{fig:bad_classification}
\end{figure}

The manual classifier highlights the primary mass as the most influential factor in distinguishing Pop.~III BBHs. However, it cannot replace \textsc{XGBoost} because it yields inadequate performance with imbalanced classes, as shown in Fig.~\ref{fig:bad_classification}. The precision with $p(j\in k | d_j, \{\beta\}) > 0.9$ evaluated with the manual classifier on the fiducial model is equal to 0.16. The performance of the manual classifier in unbalanced cases might be improved by introducing additional complexity, such as incorporating a sigmoid function and fitting its free parameters. However, in our view, this approach is more intricate than simply employing an easy-to-train ML algorithm.

\subsection{Bayes factors}

We compared our results with the method described by \cite{mould2023}, who computed the Bayes factors of single detected events generated from two different simulated populations. Specifically, we evaluated
\begin{equation}
\label{eq:bayes_dect}
    \mathcal{D}_{\mathrm{III}/\mathrm{I-II}} = \frac{P(\mathrm{det}|\mathrm{I-II})}{P(\mathrm{det}|\mathrm{III})}\mathcal{B}_{\mathrm{III}/\mathrm{I-II}}\frac{\pi(\mathrm{III})}{\pi(\mathrm{I-II})},
\end{equation}
where the ratio $\pi(\mathrm{III})/\pi(\mathrm{I-II}) = \beta_{\mathrm{III}}/\beta_{\mathrm{I-II}}$ is the prior odds ratio as chosen in Eq.~\ref{eq:rebalancing}. We included the detection probabilities $P(\mathrm{det}|\mathrm{III}) \approx 0.55 $ and $P(\mathrm{det}|\mathrm{I-II}) \approx 0.43$, which represent the fractions of detectable sources with the ET of Pop.~III and Pop.~I-II BBHs, respectively \citep[see e.g.,][]{thrane2019,mandel2019,vitale2022}. We approximated the Bayes factor between Pop.~III and Pop.~I-II BBHs $\mathcal{B}_{\mathrm{III}/\mathrm{I-II}}$ using a Monte Carlo summation,
\begin{equation}
    \mathcal{B}_{\mathrm{III}/\mathrm{I-II}} = \frac{\sum_{i = 1}^{N_s} \pi(\mathbf{x}_i|\mathrm{III})/\pi(\mathbf{x}_i|U)}{\sum_{i = 1}^{N_s} \pi(\mathbf{x}_i|\mathrm{I-II})/\pi(\mathbf{x}_i|U)},
\end{equation}
where $i$ spans $N_s$ posterior samples of a detected event $j$ $[\mathbf{x}_i \sim p(\mathbf{x}|d_j)]$, and $\pi(\mathbf{x}_i|U)$ is the probability density of the uninformative priors used during parameter estimation (see Sect.~\ref{sec:pe} and Table~\ref{tab:prior}); $\pi(\mathbf{x}|\mathrm{III})$ and $\pi(\mathbf{x}|\mathrm{I-II})$ are the astrophysical distributions of Pop. III and Pop. I-II BBH parameters, respectively (Fig.~\ref{fig:fid}). To evaluate these probability densities, we used the Gaussian kernel density estimation (\textsc{KDE}) from \textsc{sklearn} \citep{scikit-learn}, and the bandwidth was determined using the method by \cite{scott2015}.

We compared the Bayes factors in  Eq.~\ref{eq:bayes_dect} with those obtained from Eq.~\ref{eq:theequation},
\begin{equation}
\label{eq:xgbayes}
    \mathcal{X}_{\mathrm{III}/\mathrm{I-II}} = \frac{p(j \in k |d_j, \{\beta\})}{p(j \notin k |d_j, \{\beta\})} = \frac{p(j \in k |d_j, \{\beta\})}{1 - p(j \in k |d_j, \{\beta\})},
\end{equation}
where the second equality follows from the fact that $p(j \notin k |\mathbf{x}, d_j, \{\beta\}) = 1- p(j \in k |\mathbf{x}, d_j, \{\beta\})$ (see Eq.~\ref{eq:rebalancing}). Selection effects, represented by the factor $P(\mathrm{det}|\mathrm{I-II})/P(\mathrm{det}|\mathrm{III})$ in Eq.~\ref{eq:bayes_dect} are inherently accounted for in Eq.~\ref{eq:xgbayes} because the ML classifiers are trained exclusively on detectable sources. To compare the Bayes factors, we evaluated the precision metric from Eq.~\ref{eq:precision}. We placed different thresholds on the logarithms of Bayes factors to classify the same events of the fiducial model as in Fig.~\ref{fig:prob}. Table \ref{tab:bayes_ratio} shows that the two methodologies perform similarly overall, as we would expect because they ultimately approximate the same quantity with different numerical approaches. However, the classification using decision trees inherently scales better with higher-dimension data, automatically handles nonstandardized data, and is generally a factor of $\sim 10$ faster to optimize and evaluate than \textsc{KDE}.

\begin{table}
    \caption{Precision (Prec.) of the fiducial model for different thresholds (Thr.) on the logarithms of the Bayes factors obtained with \textsc{KDE} ($\ln \mathcal{D}_{\mathrm{III}/\mathrm{I-II}}$, Eq.~\ref{eq:bayes_dect}) and with \textsc{XGBoost} ($\ln \mathcal{X}_{\mathrm{III}/\mathrm{I-II}}$, Eq.~\ref{eq:xgbayes}), respectively.}
    \label{tab:bayes_ratio}
    \centering
    \begin{tabular}{l  c  c}
    \hline\hline  
        Thr. & Prec. with $\ln \mathcal{D}_{\mathrm{III}/\mathrm{I-II}}$ & Prec. with $\ln \mathcal{X}_{\mathrm{III}/\mathrm{I-II}}$  \\
        \hline
        $-3$ & 0.18 & 0.19 \\
        $-2$ & 0.24 & 0.26 \\
        $-1$ & 0.32 & 0.35 \\
        $0$  & 0.46 & 0.56 \\
        $1$  & 0.56 & 1.00 \\
        $2$  & 0.90 & 1.00 \\
        $3$  & 1.00 & 1.00 \\
        \hline
    \end{tabular}
\end{table}

\subsection{Impact of extrinsic parameters}
\label{sec:extrinsic}

In Sect.~\ref{sec:pe} we randomly chose the extrinsic parameters for each injected source based on their prior distributions (see Table~\ref{tab:prior}). These parameters, such as sky position  
and inclination angle, can affect our results. The sensitivity of the detector varies across the sky, and the luminosity distance is degenerate with the inclination angle. Consequently, altering these extrinsic parameters can either reduce or increase the total number of detected sources and the accuracy of their parameter estimations.

We generated ten catalogs drawing different independent values for the extrinsic parameters. This procedure was performed exclusively for the pessimistic model because it is more likely to be sensitive to the choice of extrinsic parameters: It has fewer expected mergers than the other models. Subsequently, we injected these new populations in \textsc{GWFish}.

When we varied the extrinsic parameters ten times while keeping the intrinsic parameters fixed (i.e., $m_{1,\mathrm{d}}$, $m_{2,\mathrm{d}}$, and $d_{\mathrm{L}}$), the total number of detected sources with a relative error of $\leq 0.3$ fluctuates between  $N_{\mathrm{Pop.\ III}} = 13$ and $N_{\mathrm{Pop.\ III}} = 25$. The percentage of BBHs correctly identified with $ p(j\in k | d_j, \{\beta\}) \geq 0.1$ as being generated from Pop.~III ranges between 33\% and 75\% with a corresponding precision between 0.4 and 1.

\subsection{Spin parameters}
\label{sec:spins}

In our analysis, we set spin parameters equal to zero. This is because \textsc{sevn} only provides as output the tilt angle in the setup adopted by \cite{costa2023}. This is the angle between the rotation axis of the single BH with the direction of the angular momentum of the orbital system \citep{iorio2023}. There is no significant difference in the distribution of cosines of tilt angles between Pop.~I-II and Pop.~III BBHs because $>99$\% of them are equal to one in both populations and across redshift. In our case, the spin magnitudes must be added in post-processing \citep[e.g.,][]{Wysocki2018,gerosa2018,Belczynski2020,bavera2020,perigois2023}, which means that the differences, if any, between Pop.~III and Pop.~I-II BBHs are artificial.

\subsection{Fisher information-matrix approximation}
\label{sec:caveatsonFIM}

The FIM formalism is widely used in forecasting studies, where it shapes the design and scientific objectives of advanced detectors \citep{Borhanian:2022czq,pieroni2022,iacovelli2022b,branchesi2023}. However, this approximation is only valid under certain conditions, most notably, the high-S/N limit \citep{vallisneri2008}. Even for loud sources, disparities may emerge between the parameter uncertainties estimated using the FIM formalism and those derived through full Bayesian inference \citep{vandersluys,rodriguez2013,Mandel2014,veitch2015,bilby}. Nonetheless, the chosen priors in Table~\ref{tab:prior} and the cuts on relative errors and S/N mitigate this effect \citep{Dupletsa2024}.

\subsection{Population modeling uncertainty}
\label{sec:hyperunc}

The general performances of the classifiers may be influenced by the interplay of three factors. First, the simulated populations of Pop.~I-II and Pop.~III BBHs might differ from those that will be observed with the ET, potentially leading to biased predictions. Most existing machine-learning models operate under the closed-world assumption, assuming that the test data are drawn from the same distribution as the training data. However, in open-world scenarios, events can be unknown and deviate significantly from the training data. Therefore, it is essential to ensure that the simulated data accurately reflect reality. Several out-of-distribution detection techniques have been proposed, ranging from classification-based to density-based to distance-based methods \citep[see][for a review]{yang2022generalized}. One such technique was presented in \cite{pasquato2023}, who estimated the probability density of the training set and evaluated it on both the test set and real dataset. They used KDE scores to identify and reject out-of-distribution data points. Similarly, we will apply our method only once it has been determined that the simulated data matches reality. Additionally, retraining on alternative data is easily achievable because \textsc{XGBoost} requires only few minutes on ten CPUs for training, considering both optimization and cross-validation.

The classification of detected events can also face challenges by sources that originate from different astrophysical phenomena. For instance, the dynamical formation channels predict a subpopulation of BBHs with primary masses ranging from 40 to 100~M$_\odot$ \citep[e.g.,][]{ziosi2014, 
 Rodriguez2015, Rodriguez2019, antonini2016,mapelli2016,askar2017,  Banerjee2017,Banerjee2021,antonini2018,dicarlo2019,dicarlo2020,arcasedda2019,rastello2020,kumamoto2020,arcasedda2023,kritos2023,arcasedda2024}. When we consider the peak of their merger rate density at redshift $z \sim 2$ \citep{santoliquido2020}, dynamically formed BHs would exhibit a detector-frame primary mass in the range of $\ssim$80 to 200~M$_\odot$. Consequently, the tails of the distribution for Pop.~I-II BBHs would be thicker in the overlapping regions with Pop.~III BBHs, potentially leading to less accurate classification performances (see Figs.~\ref{fig:fid}, \ref{fig:opt} and \ref{fig:pes}). Moreover, Pop.~III BBHs could be subject to contamination from primordial black holes, which are expected to form at high redshift with a wide range of masses \citep[e.g.,][]{deluca2020,deluca22,franciolini2022,ng2022}. When we consider the possibility of sources with various astrophysical origins, which we leave for future studies, it might become crucial to account for additional features to effectively disentangle them, for instance, spin properties \citep{gerosa2013,qin2018,qin2019,fuller2019,bavera2020,Belczynski2020,Olejak2021,stevenson2022,perigois2023} and eccentricity \citep{Samsing2014,samsing2018,zevin2019,arcasedda2021,Romero-Shaw2023,codazzo2023,dallamico2023}. Additionally, a network of third-generation detectors might be considered to enhance the precision in the parameter estimation \citep{iacovelli2022, Borhanian:2022czq}. 

In our analysis, while evaluating $p(j\in k | d_j, \{\beta\})$ from \eqref{eq:theequation}, we assumed that the hyperparameters of the population governing the distribution of single-event parameters are perfectly known. This is not generally true even with the ET, where uncertainties on hyperparameters are expected. Posterior distributions of population hyperparameters can be inferred by considering a set of gravitational-wave observations via a hierarchical Bayesian analysis \citep[e.g.][]{loredo2004,mandel2019,bouffanais2021a,zevin2021,vitale2022}.
We leave for future studies the integration of the hyperparameter uncertainty on $p(j\in k | d_j, \{\beta\})$ and the inclusion of population-informed priors \citep{miller2020,moore2021,abbottO3bpopandrate} to analyze the single-event posterior 
($p({\bf{x}}|d_j,\{\beta\})$ in \eqref{eq:theequation}). In this way, we will quantify the variance that affects our results due to modeling systematics.

\section{Conclusions}
\label{sec:conclusion}

The ability to trace the origins of individual GW events upon their detection is pivotal for understanding the formation and evolution of compact objects. To address this problem, and with the goal of enhancing the scientific output of future GW astronomy, we coupled the heightened sensitivity of the ET with innovative ML techniques. We proposed a robust and straightforward ML approach that leverages the power of \textsc{XGBoost} \citep{xgboost}, a fast-to-train algorithm based on decision trees that leads to very good performances.

\citet{costa2023} investigated a broad parameter space of initial conditions of Pop.~III progenitor stars in binary systems using advanced population-synthesis simulations, based on stellar tracks   \citep[\textsc{sevn},][]{iorio2023}. \cite{santoliquido2023} combined these Pop.~III BBHs  with different star formation rate histories to estimate their merger density evolution with redshift \citep[\textsc{cosmo$\mathcal{R}$ate},][]{santoliquido2021}. From the array of available models, we specifically chose three populations of Pop.~III BBHs (fiducial, optimistic, and pessimistic) based on their respective detection rates. 
They were classified against a population of Pop.~I-II stars taken from the fiducial model by \cite{iorio2023}. The parameter estimation of these populations was performed using the FIM formalism  \citep[\textsc{GWFish},][]{dupletsa2023}. 

We  trained \textsc{XGBoost} on balanced datasets, which yielded very good performances with a precision exceeding 90\% for all the models (Sect.~\ref{sec:classifier}). We applied these classifiers in an event-by-event scenario, in which posterior samples of detected sources, the relative ratio of Pop.~III and Pop.~I-II BBH merger rate, and classification confidence via $p(j \in k| d_j, \{\beta\})$ were taken into account (Sect.~\ref{sec:post_samples}). Our analysis revealed that the classifiers consistently achieve a precision of $>90\%$ in classifying a fraction of BBHs from Pop.~III stars. In particular, we confidently identified $\gtrsim 10\%$, $\gtrsim 50\%$, and $\lesssim 50\%$ of BBHs from Pop.~III stars in the fiducial, optimistic, and pessimistic  models, respectively, with a precision of $>90\%$ (Table~\ref{tab:fid}). 

This study proposes a new method that integrates the simulation-based information of GW source populations with the parameter estimation inference. We demonstrated its validity in discerning the origins of individual GW detections.

\begin{acknowledgements}
We acknowledge the anonymous referee for their valuable comments. We thank Biswajit Banerjee, Manuel Arca Sedda, Samuele Ronchini, Stefano Torniamenti, Elena Cuoco, Guglielmo Costa, Davide Piras and Franco Raimondi for valuable discussions. This work was performed in part at the Aspen Center for Physics, which is supported by National Science Foundation grant PHY-2210452 and by a grant from the Simons Foundation (1161654, Troyer). F.S. and M.B. acknowledge financial support from the AHEAD2020 project (grant agreement n. 871158). M.B. also acknowledges support from the PRIN grant METE under  contract No. 2020KB33TP. M.M., F.S., and G.I. acknowledge financial support from the European Research Council for the ERC Consolidator grant DEMOBLACK, under contract No. 770017 (PI: Mapelli). M.M. also acknowledges support from the German Excellence Strategy via the Heidelberg Cluster of Excellence (EXC 2181 - 390900948) STRUCTURES. D.G. is supported by
ERC Starting Grant No.~945155--GWmining,
Cariplo Foundation Grant No.~2021-0555,
MUR PRIN Grant No.~2022-Z9X4XS, 
MSCA Fellowships No.~101064542--StochRewind and No.~101149270--ProtoBH,
and the ICSC National Research Centre funded by NextGenerationEU. F.I. is supported by Swiss National Science Foundation Grant No.~200020$\_$191957 and the SwissMap National Center for Competence in Research. The research leading to these results has been conceived and developed within the \textit{Einstein} Telescope Observational Science Board ET-0101A-24. U.D. acknowledges Stefano Bagnasco, Federica Legger, Sara Vallero, and the INFN Computing Center of Turin for providing support and computational resources. G.I. acknowledges financial support under the National Recovery and Resilience Plan (NRRP), Mission 4, Component 2, Investment 1.4, - Call for tender No. 3138 of 18/12/2021 of Italian Ministry of
University and Research funded by the European Union – NextGenerationEU
\end{acknowledgements}

\section*{Data Availability}
\textsc{sevn} is publicly available at \href{https://gitlab.com
/sevncodes/sevn.git}{https://gitlab.com
/sevncodes/sevn.git}: the version used in this work is the commit
\textsc{0f9ae3bf} in the branch \textsc{Costa23popIII} (\href{https://gitlab.com/s
evncodes/sevn/-/tree/Costa23popIII}{https://gitlab.com/s
evncodes/sevn/-/tree/Costa23popIII}). \textsc{cosmo$\mathcal{R}$ate} is publicly available at \href{https://gitlab.com/filippo.santoliquido/cosmo_rate_public}{gitlab.com/Filippo.santoliquido/cosmo\_rate\_public}. \textsc{GWFish} is publicly available
 at \href{https://github.com/janosch314/GWFish}{github.com/janosch314/GWFish}. A dataset with the parameter estimation generated using \textsc{GWFish} is available on Zenodo \citep{santoliquido_2024_14013888}. The code for reproducing all results and figures presented in this paper is available at \href{https://github.com/filippo-santoliquido/ET_classifier}{https://github.com/filippo-santoliquido/ET\_classifier}.



%
%
\bibliographystyle{aa_edited.bst} 
\bibliography{main.bib} 

\end{document}